\documentclass{jfm}

\usepackage{graphicx}
\usepackage{newtxtext}
\usepackage{newtxmath}
\usepackage{natbib}
\usepackage{hyperref}
\hypersetup{
    colorlinks = true,
    urlcolor   = blue,
    citecolor  = black,
}

\usepackage{makecell}
\usepackage{placeins}
\usepackage{bm}
\usepackage{overpic}[perc]
\usepackage[mathscr]{euscript}

\graphicspath{{./plots/}}

\newcommand{\RomanNumeralCaps}[1]
\linenumbers

\newcommand{\imag}{\mathrm{i}}
\newcommand{\inv}{\mathcal{I}}
\newcommand{\vx}{\vect{x}}
\newcommand{\vk}{\bm{k}}

\newcommand{\FS}{\widetilde{S}}

\newcommand{\Fu}{\widetilde{u}}
\newcommand{\Fk}{\bm{q}}

\newcommand{\de}{\textrm{d}}
\newcommand{\Tr}{\textrm{Tr}}

\newcommand{\ordof}{\textit{O}}
\newcommand{\avg}[1]{\left\langle #1 \right\rangle}
\newcommand{\cavg}[2]{\left.\left\langle #1 \middle| #2 \right\rangle \right.}

\newcommand{\of}[1]{\left( #1 \right)}
\newcommand{\vect}[1]{\bm{#1}}
\newcommand{\tens}[1]{\mathsfbi{#1}}

\title{Asymptotic predictions on the velocity gradient statistics in low-Reynolds number random flows: Onset of skewness, intermittency and alignments}

\author{
Maurizio Carbone\aff{1,2}
  \corresp{\email{maurizio.carbone@uni-bayreuth.de}}
 \and Michael Wilczek\aff{1,2}
  \corresp{\email{michael.wilczek@uni-bayreuth.de}}}
\affiliation{\aff{1}Max Planck Institute for Dynamics and Self-Organization, Am Fa{\ss}berg 17, 37077 Göttingen, Germany
\aff{2} Theoretical Physics I,  University of Bayreuth, Universit\"atsstr.\ 30, 95447 Bayreuth, Germany}

\begin{document}
\maketitle

\begin{abstract}
Stirring a fluid through a Gaussian forcing at a vanishingly small Reynolds number produces a Gaussian random field, while flows at higher Reynolds numbers exhibit non-Gaussianity, cascades, anomalous scaling and preferential alignments.
Recent works (Yakhot and Donzis, \textit{Phys.~Rev.~Lett.}, vol.~119, 2017, pp.~044501; Gotoh and Yang, \textit{Philos.~Trans.~Royal Soc.~A}, vol.~380, 2022, pp.~20210097) investigated the onset of these turbulent hallmarks in low-Reynolds number flows by focusing on the scaling of the velocity increments. They showed that the scalings in random flows at low-Reynolds and in high-Reynolds number turbulence are surprisingly similar.
In this work, we address the onset of turbulent signatures in low-Reynolds number flows from the viewpoint of the velocity gradient dynamics, giving insights into its rich statistical geometry. We combine a perturbation theory of the full Navier-Stokes equations with velocity gradient modeling. This procedure results in a stochastic model for the velocity gradient in which the model coefficients follow directly from the Navier-Stokes equations and statistical homogeneity constraints.
The Fokker-Planck equation associated with our stochastic model admits an analytic solution which shows the onset of turbulent hallmarks at low Reynolds numbers: skewness, intermittency and preferential alignments arise in the velocity gradient statistics as the Reynolds number increases.
The model predictions are in excellent agreement with direct numerical simulations of low-Reynolds number flows.
\end{abstract}


\section{Introduction}
\label{sec_introduction}

The Reynolds number characterizes the range of active scales involved in the flow of a Newtonian fluid, determining the transition from a laminar motion to a disordered and turbulent state \citep{Reynolds1883}.
At a vanishingly small Reynolds number, the flow obeys linear equations and stirring the fluid through a Gaussian random forcing produces a Gaussian random velocity field. On the other hand, flows at higher Reynolds numbers undergo a non-linear evolution and exhibit highly non-Gaussian turbulent features. These distinguishing marks of the turbulent dynamics include
cascades \citep[e.g.][]{Alexakis2018,Ballouz2020,Vela-martin2021}, anomalous scaling of the velocity increments \citep[e.g.][]{Benzi1995,Chen2005},
extreme intermittency and preferential alignments of the velocity gradients \citep[e.g.][]{Ashurst1987,Lund1994,Meneveau2011,Buaria2019,Buaria2022}.
Recent works \citep{Yakhot2017,Sreenivasan2021,Gotoh2022,Khurshid2022} characterized the onset of these turbulent features and highlighted a striking similarity between the scalings in low-Reynolds number flows and fully developed turbulence, which motivates further investigations of low-Reynolds number random flows.

One approach to address the onset of turbulent motion consists of considering the three-dimensional, incompressible Navier-Stokes equations for a statistically isotropic flow without boundaries, driven by a large-scale Gaussian forcing.
This setup excludes the effect of any boundary condition and partially overcomes the lack of universality of low-Reynolds number flows \citep{Gotoh2022}.
Such idealized flows can be investigated analytically at a small Reynolds number by formulating a perturbation theory of the Navier-Stokes equations \citep{Wyld1961}. This direct approach provides complete insight into the full velocity field at low Reynolds numbers but involves several technical complications.
For example, the terms in the series expansion soon become excessively complicated and violations of Galilean invariance occur \citep{Yakhot2018}.

A recent insightful approach focused on the scaling of the velocity gradient moments and structure functions at low Reynolds number \citep{Yakhot2017,Sreenivasan2021} starting from the Hopf equation for the characteristic functional of the velocity field \citep{Hopf1952}. This scaling analysis showed that, surprisingly, low-Reynolds number flows without an inertial range, not detected even with the aid of the extended self-similarity \citep{Benzi1993}, have qualitatively the same scalings as fully developed turbulence \citep{Schumacher2007}.
In particular, the scaling exponents of the structure functions, with the Reynolds number and spatial separation, observed in low-Reynolds number flows match well the scaling exponents predicted by a theory relying on very high Reynolds number hypotheses \citep{Yakhot2004}. Those scalings are observed at spatial separations $r\gtrapprox\eta$, where $\eta$ is the Kolmogorov length scale, and at a Reynolds number based on the Taylor micro-scale $\Rey_\lambda\gtrapprox 9$, whereas any anomalous scaling is negligible at $\Rey_\lambda\le 3$ \citep{Yakhot2017}. While this scaling analysis fully characterizes the velocity increment statistics across the scales, it does not shed light on the rich statistical geometry of the flow. For example, the alignments and interplay between the strain rate and the vorticity cannot be inferred.

Here, we address the onset of non-Gaussianity in flows driven by a random forcing from the viewpoint of the velocity gradients.
The velocity gradient encodes many distinguishing features of the turbulent state \citep{Meneveau2011}, and it comprehensively characterizes the geometry of the vorticity and strain rate.
As the Reynolds number increases, the velocity gradient transitions from a Gaussian random matrix state \citep{Livan2018} to a turbulent state, featuring skewness of the longitudinal components, associated with the cascade of kinetic energy \citep{Eyink2006,Carbone2020a,Johnson2020},
and preferential configurations of the strain-rate eigenvalues \citep{Betchov1956,Lund1994,Davidson2015}, as well as intermittency and preferential alignments between the vorticity and the strain-rate eigenvectors \citep{Ashurst1987,Buaria2020}.

To analytically capture the onset of non-Gaussianity, we construct a model for the velocity gradients that is directly derived from the randomly-forced Navier-Stokes equations at low Reynolds numbers.
The main challenge in formulating such a model for the gradient dynamics stems from the non-locality of turbulence. The drastic reduction of degrees of freedom in going from the full Navier-Stokes equations to a small system of ordinary differential equations governing the gradient dynamics comes at the cost of introducing unclosed terms \citep{Meneveau2011}.
Those unclosed terms consist of the traceless/anisotropic pressure Hessian and the viscous Laplacian of the velocity gradient, which require modeling. 
Recent phenomenological models for the velocity gradient have proven effective in reproducing the small-scale turbulence statistics at moderately large Reynolds numbers  \citep[e.g.][]{Girimaji1990a,Chertkov1999,Chevillard2006,Wilczek2014,Johnson2016,Pereira2018,Leppin2020}.

In contrast to phenomenological models at high Reynolds numbers, our low-Reynolds number model for the velocity gradients can be derived directly from the Navier-Stokes equations. This direct derivation reduces the number of modelling hypotheses and free parameters. Indeed, at order zero in the Reynolds number, the velocity field is Gaussian, allowing for the exact computation of the non-local/unclosed terms in the equations governing the velocity gradient \citep{Wilczek2014}.
We use those exact expressions of the unclosed terms at zero Reynolds number to construct an expansion in the Reynolds number of the velocity gradient dynamics.
Then, we close the model by using the asymptotic weak-coupling expansion of the full Navier-Stokes equations at small Reynolds number \citep{Wyld1961}, combined with the two statistical homogeneity constraints on the incompressible velocity gradient \citep{Betchov1956,Carbone2022}.
The resulting model for the single-time statistics of the velocity gradient does not feature adjustable parameters and does not require any input from simulations or experiments.
Furthermore, we extend the model to predict the full temporal dynamics by using two adjustable model parameters fitted from DNS results. This extended model captures both the velocity gradient single-time statistics and the time correlations.

The presented model is associated with a Fokker-Planck equation for the velocity gradient probability density function (PDF), in which the Reynolds number and forcing parameters are in one-to-one correspondence with the same parameters featured in the forced Navier-Stokes equations. This Fokker-Planck equation admits asymptotic analytic solutions, thus yielding an analytic approximation to the velocity gradient PDF.

Attempts to analytically predict the high-dimensional velocity gradient PDF are limited and, so far, built upon phenomenological models for the small-scale turbulent dynamics \citep{Chertkov1999,Moriconi2014,Apolinario2019}. In general, the analytical expression for the PDF follows from field-theoretical methods employed to solve the non-linear Langevin equation governing the flow dynamics \citep{Martin1973}, consisting of the renormalized action method with a one-loop correction \citep{Kleinert2001,Cavagna2021}. The resulting form of the PDF of the gradient is typically not integrable analytically, thus preventing the direct computation of marginal distributions and moments, computed instead via Monte-Carlo sampling \citep{Moriconi2014}. 
Differently, our prediction of the PDF of the velocity gradient is analytically integrable and it allows us to derive expressions for the marginal distributions and the moments of the velocity gradients. Those analytic expressions explicitly relate the quantities of interest, e.g.~skewness, kurtosis and preferential alignments of the gradients, to the Reynolds number and forcing parameters, thus rationalizing the onset of non-Gaussianity in low-Reynolds random flows.

The paper is organized as follows. Section \ref{sec_theory} presents the derivation of a Fokker-Planck equation (FPE) for the velocity gradient PDF from the Navier-Stokes equations. Section \ref{sec_model} specializes the FPE for low-Reynolds flows, while section \ref{sec_solution} presents its analytic solution. The comparison between the model/analytic predictions and low-Reynolds DNS is presented in section \ref{sec_results}, while the conclusions and outlook are discussed in section \ref{sec_conclusions}. The appendices \ref{app_DNS_details} and \ref{app_Wyld} describe the setup of the numerical simulations and the low-Reynolds number expansion of the Navier-Stokes equations used to determine the model coefficients.
\section{Fokker-Planck equation for the velocity gradient probability density in statistically isotropic flows}
\label{sec_theory}

In this section, we obtain the general Fokker-Planck equation (FPE) governing the single-time/single-point statistics of the velocity gradient. We then specialize it for flows at low Reynolds numbers.

\subsection{Equation governing the velocity gradient dynamics}

We begin with the three-dimensional incompressible Navier-Stokes equations, driven by an external Gaussian stochastic forcing $\vect{F}$,
\begin{subequations}
\begin{align}
\bnabla\bcdot\vect{u} &= 0, \\
\partial_t \vect{u} + \Rey_\gamma\left[(\vect{u} \bcdot \bnabla) \vect{u} + \bnabla P \right] &= \nabla^2 \vect{u} + \sigma \vect{F},
\label{eq_NS_ndim_mom}
\end{align}
\label{eq_NS_ndim}
\end{subequations}
where $\vect{u}(\vect{x},t)$ is the velocity field, $P(\vect{x},t)$ is the pressure field and $\Rey_\gamma$ is the Reynolds number.
Equations \eqref{eq_NS_ndim} are written in non-dimensional variables suited for the upcoming low-Reynolds number expansion.
We will employ non-dimensional variables throughout the paper, and denote the corresponding dimensional variables with a bar when needed.
To construct reference scales for a suitable non-dimensionalization of the variables, we introduce a reference length $\bar{\gamma}_0$, related to the  damping role of the viscous Laplacian at low Reynolds numbers (see \eqref{eq_V_Gauss} for its definition).
With $\bar{\gamma}_0$, together with the kinematic viscosity of the fluid $\bar{\nu}$ and the Kolmogorov time scale $\bar{\tau}_\eta=1/\sqrt{\langle \|\bar{\bnabla}\bar{\vect{u}}\|^2\rangle}$ (angle brackets indicating ensemble average), we define the non-dimensional variables employed in equation \eqref{eq_NS_ndim} as
\begin{align}
t = \left(\bar{\nu}/\bar{\gamma}_0^2\right)\bar{t}, &&
\vect{x} = \bar{\vect{x}}/\bar{\gamma}_0, &&
\vect{u} = \left(\bar{\tau}_\eta/\bar{\gamma}_0\right)\bar{\vect{u}}, &&
\sigma^2 = \left(\bar{\gamma}_0^2\bar{\tau}_\eta^2/\bar{\nu}\right)\bar{\sigma}^2.
\label{def_nd_variables}
\end{align}
Based on these quantities, we can define a Reynolds number according to
\begin{align}
\Rey_\gamma = \frac{\bar{\gamma}_0^2}{\bar{\nu}\bar{\tau}_\eta},
\label{def_Re_gamma}
\end{align}
which expresses the ratio between the velocity gradient magnitude $\bar{\tau}_\eta^{-1}$ and the viscous damping $\bar{\nu}/\bar{\gamma}_0^2$.
The Reynolds number \eqref{def_Re_gamma} weighs the nonlinearities in equation \eqref{eq_NS_ndim}, thus allowing to take the small-Reynolds-number limit properly.
The zeroth-order solution of the non-dimensional Navier-Stokes equation \eqref{eq_NS_ndim} consists of a Gaussian random velocity field resulting from a stochastically-driven diffusion equation. By gradually increasing the Reynolds number, the nonlinear terms come into play leading to non-Gaussian statistics. In the following, we analyze this ``onset of non-Gaussianity" at low Reynolds numbers.

The stochastically-driven Navier-Stokes equations \eqref{eq_NS_ndim} are associated with a Fokker-Planck equation for the velocity gradient single-time/single-point probability density \citep{Wilczek2014}.
To obtain it, we start with the Langevin equation governing the velocity gradient dynamics,
obtained from taking the gradient of equation \eqref{eq_NS_ndim_mom},
\begin{subequations}
\begin{align}
\Tr(\tens{A}) &= 0, \\
\partial_t \tens{A}  + \Rey_\gamma\vect{u}\bcdot\bnabla\tens{A} &= - \Rey_\gamma\left(\tens{AA} + \tens{H} \right) + \nabla^2 \tens{A} + \sigma \bnabla \vect{F},
\end{align}
\label{eq_Langevin_A}
\end{subequations}
where $A_{ij} = \nabla_j u_i$ is the velocity gradient, $H_{ij} = \nabla_i\nabla_j P$ is the pressure Hessian and standard matrix product is implied.
The Gaussian tensorial noise $\bnabla \vect{F}$ in \eqref{eq_Langevin_A} has zero mean, is statistically isotropic and white in time. It is fully specified by its correlation, which in Cartesian component notation reads
\begin{align}
\avg{\nabla_j F_i (\vect{x},t) \nabla_q F_p(\vect{x},t')} = \delta(t-t')\left( 4\delta_{ip}\delta_{jq}-\delta_{iq}\delta_{jp}-\delta_{ij}\delta_{pq} \right),
\label{eq_f_correl_phys}
\end{align}
where $\delta_{ij}$ denotes the Kronecker delta.
The spatial correlation of the stochastic forcing is specified in Appendix \ref{app_DNS_details}.

\subsection{Fokker-Planck equation for the velocity gradient invariants}

Equation \eqref{eq_Langevin_A} is associated with a Fokker-Planck equation (FPE) for the probability density function (PDF) of the velocity gradient $f(\tens{A}; t)$. In Cartesian components, the FPE for an ensemble of fluid particles sharing the same instantaneous configuration of the velocity gradient $\tens{A}$ reads \citep{Wilczek2014},
\begin{align}
\frac{\partial f}{\partial t} &=
\frac{\partial}{\partial A_{ij}} \Bigg[ \Rey_\gamma\left(A_{ik}A_{kj} + \left\langle H_{ij}(\vect{x},t)|\tens{A}\right\rangle\right)f - \left\langle\nabla^2 A_{ij}(\vect{x},t)|\tens{A}\right\rangle f + \nonumber\\
&+\frac{1}{2}\sigma^2\left(4\frac{\partial f}{\partial A_{ij}}  - \frac{\partial f}{\partial A_{ji}} \right)\Bigg],
\label{eq_FPE_Cart}
\end{align}
where $\langle\cdot |\tens{A} \rangle$ denotes the ensemble average conditional on the velocity gradient configuration $\tens{A}$.
The conditional averages in \eqref{eq_FPE_Cart}, namely the conditional anisotropic pressure Hessian and viscous Laplacian of the gradient, are unclosed terms that cannot be computed based only on the gradient at a single point \citep{Meneveau2011}. Thus, the simplifications of going from the equation for the whole velocity gradient \eqref{eq_Langevin_A} (encoding the full space-time complexity of the velocity gradient field realizations), to an equation for the single-time/single-point statistics of the gradients \eqref{eq_FPE_Cart}, come with the cost of introducing unclosed terms.

In statistically isotropic flows, the velocity gradient probability density function $f$ is rotationally invariant, namely a function of only the five independent velocity gradient invariants \citep{Itskov2015}.
Statistical isotropy then allows us to employ tensor function representation theory  \citep[e.g.][]{Rivlin1955,Itskov2015} to express the unclosed conditional averages in \eqref{eq_FPE_Cart} as isotropic tensor functions of the velocity gradient.
The conditional averages are represented as combinations of basis tensors $\tens{B}^n$ with coefficients $\gamma_n$ that depend upon the velocity gradient invariants $\inv_k$.
More specifically, the basis tensors are formed through the strain-rate tensor $\tens{S}=(\tens{A} + \tens{A}^\top)/2$ and rotation-rate tensor $\tens{W}=(\tens{A} - \tens{A}^\top)/2$, and they read
\begin{align}
\begin{aligned}[c]
\tens{B}^1 &= \tens{S} \\
\tens{B}^2 &= \tens{W}
\end{aligned}
&&
\begin{aligned}[c]
\tens{B}^3  &= \widetilde{\tens{SS}} \\
\tens{B}^4  &= \tens{S} \tens{W} - \tens{W} \tens{S}\\
\end{aligned}
&&
\begin{aligned}[c]
\tens{B}^5  &= \tens{S} \tens{W} + \tens{W} \tens{S}\\
\tens{B}^6 &= \widetilde{\tens{WW}} \\
\end{aligned}
&&
\begin{aligned}[c]
\tens{B}^7 &= \widetilde{\tens{S} \tens{WW}} + \widetilde{\tens{WW} \tens{S}}  \\
\tens{B}^8 &= \widetilde{\tens{SS} \tens{W}} + \widetilde{\tens{W} \tens{SS}},
\end{aligned}
\label{eq_def_BT}
\end{align}
where the tilde indicates the traceless/anisotropic part of the tensor, and standard matrix product is implied. It would be necessary to include two additional basis tensors in \eqref{eq_def_BT} to fix a possible degeneracy of the basis \citep{Pennisi1987}. Indeed, some of the tensors \eqref{eq_def_BT} become linearly dependent when two of the strain-rate eigenvalues coincide or when the vorticity is an eigenvector of the strain rate. We neglect those zero-measure configurations.
The independent invariants $\inv_k$ formed through the velocity gradient read
\begin{align}
\inv_1 &= \Tr\left(\tens{S} \tens{S}\right) &&
\inv_3 = \Tr\left(\tens{S} \tens{S} \tens{S}\right) &&
\inv_5 = \Tr\left(\tens{S} \tens{S} \tens{W} \tens{W}\right),
\nonumber\\
\inv_2 &= \Tr\left(\tens{W} \tens{W}\right) &&
\inv_4 = \Tr\left(\tens{S} \tens{W} \tens{W}\right).
\label{eq_def_inv}
\end{align}
A sixth invariant would be necessary to fix the handedness of the strain-rate eigenvector basis with respect to the vorticity. However, this sixth invariant is uniquely determined in terms of the other five only up to a sign \citep{Lund1992} and we do not consider it as an independent variable.

With the above definitions of the basis tensors \eqref{eq_def_BT} and invariants \eqref{eq_def_inv}, the drift term in the FPE \eqref{eq_FPE_Cart} can be compactly written as
\begin{align}
\Rey_\gamma\left(\tens{AA} + \left\langle \tens{H}(\vect{x},t)\middle|\tens{A}\right\rangle\right) - 
\left\langle\nabla^2 \tens{A}(\vect{x},t)\middle|\tens{A}\right\rangle
 = \sum_{n=1}^{8} \gamma_n(\inv) \tens{B}^n.
\label{eq_tens_repr}
\end{align}
The general expression \eqref{eq_tens_repr} allows us to formulate the FPE \eqref{eq_FPE_Cart} in terms of the invariants by carrying out the contractions using symbolic calculus  \citep{Sympy2017}.
For notation simplicity, we denote the PDF of the velocity gradient with $f$ regardless of its argument, $f(\inv(\tens{A})) \equiv f(\tens{A})$ since $f$ is a probability density with respect to $\tens{A}$ and the invariants $\inv_k$ are employed only for a simpler (isotropic) parametrization.
Finally, the steady-state FPE \eqref{eq_FPE_Cart} reduces to
\begin{align}
\gamma_n \phi^n f &+ M_{kl}Z^{ln} \frac{\partial \left(\gamma_n f\right)}{\partial \inv_k} -
\frac{\sigma^2}{2} M_{kl}\of{4\phi^l-\phi'^l}\frac{\partial  f}{\partial \inv_k} =\nonumber\\
&=
\frac{\sigma^2}{2} M_{kl}M_{pq}\of{4Z^{lq}-Z'^{lq}}\frac{\partial^2 f}{\partial \inv_k \partial \inv_{p}} ,
\label{eq_FPE_inv}
\end{align}
with sum over repeated indices implied.
The symmetric matrix $\tens{Z}$ in equation \eqref{eq_FPE_inv} is the metric tensor $Z^{ln}(\inv) = B^l_{ij}B^n_{ij}$, which in matrix form reads
\begin{align}
\tens{Z} = 
&\tiny{
\left[\begin{matrix}\inv_1 & 0 & \inv_3 & 0 & 0 & \inv_4 & 2 \inv_5 & 0\\0 & - \inv_2 & 0 & 0 & - 2 \inv_4 & 0 & 0 & - 2 \inv_5\\\inv_3 & 0 & \frac{\inv_1^{2}}{6} & 0 & 0 & - \frac{\inv_1 \inv_2}{3} + \inv_5 & \frac{\inv_1 \inv_4}{3} + \frac{2 \inv_2 \inv_3}{3} & 0\\0 & 0 & 0 & \inv_1 \inv_2 - 6 \inv_5 & 0 & 0 & 0 & 0\\0 & - 2 \inv_4 & 0 & 0 & - \inv_1 \inv_2 + 2 \inv_5 & 0 & 0 & - \inv_1 \inv_4 - \frac{\inv_2 \inv_3}{3}\\\inv_4 & 0 & - \frac{\inv_1 \inv_2}{3} + \inv_5 & 0 & 0 & \frac{\inv_2^{2}}{6} & \frac{\inv_2 \inv_4}{3} & 0\\2 \inv_5 & 0 & \frac{\inv_1 \inv_4}{3} + \frac{2 \inv_2 \inv_3}{3} & 0 & 0 & \frac{\inv_2 \inv_4}{3} & - \frac{\inv_1 \inv_2^{2}}{2} + 3 \inv_2 \inv_5 + \frac{2 \inv_4^{2}}{3} & 0\\0 & - 2 \inv_5 & 0 & 0 & - \inv_1 \inv_4 - \frac{\inv_2 \inv_3}{3} & 0 & 0 & \frac{\inv_1^{2} \inv_2}{2} - 3 \inv_1 \inv_5 + \frac{2 \inv_3 \inv_4}{3}\end{matrix}\right]}
\label{eq_def_met_tens}
\end{align}
while $Z'^{ln} = B^l_{ij}B^n_{ji}$ denotes a modified metric tensor.
The symbols $\phi^n(\inv) = \partial B^n_{ij}/\partial A_{ij}$ indicate the divergence of the basis tensors
\begin{align}
\vect{\phi} =
\left[\begin{matrix}5 & 3 & 0 & 0 & 0 & 0 & \frac{10 \inv_2}{3} & 2 \inv_1\end{matrix}\right],
\label{eq_def_div_BT}
\end{align}
while the divergence of the transposed basis tensors is $\phi'^n = \partial B^n_{ji}/\partial A_{ij}$.
The components of the derivatives of the invariants are collected in the matrix $M_{kl}=(\partial \inv_k/\partial A_{ij})B^l_{ij}Z^{-1}_{nl}$ (with $\tens{Z}^{-1}$ denoting the inverse of the metric tensor \eqref{eq_def_met_tens}), which in matrix form reads
\begin{align}
\tens{M} = 
\left[\begin{matrix}2 & 0 & 0 & 0 & 0 & 0 & 0 & 0\\0 & -2 & 0 & 0 & 0 & 0 & 0 & 0\\0 & 0 & 3 & 0 & 0 & 0 & 0 & 0\\0 & 0 & 0 & 0 & -1 & 1 & 0 & 0\\0 & 0 & 0 & 0 & 0 & 0 & 1 & -1\end{matrix}\right].
\label{eq_def_dinv_comp}
\end{align}
The quantities (\ref{eq_def_met_tens}, \ref{eq_def_div_BT}, \ref{eq_def_dinv_comp}) characterizing the tensor basis \eqref{eq_def_BT} can all be derived from the Christoffel symbols computed in \cite{Carbone2022}.

Equation \eqref{eq_FPE_inv} is a second-order partial differential equation for the function $f(\inv)$ of the five variables $\inv_k$. The model coefficients $\gamma_n(\inv)$ determine the properties and complexity of the FPE, and we will now compute the model coefficients $\gamma_n$ for low-Reynolds number flows.

\section{Determining the model coefficients}
\label{sec_model}

We introduce the two hypotheses necessary to obtain the presented model:
the coefficients $\gamma_n$ in \eqref{eq_tens_repr} are constant and the expansion \eqref{eq_tens_repr} is truncated to basis tensors up to degree two in the velocity gradient.
While these assumptions are exact for the conditional pressure Hessian at first order in Reynolds number, they introduce modelling approximations for the viscous stresses. We will test the validity of the assumptions a-posteriori, by comparing the model coefficients $\gamma_n$ with the same coefficients computed directly from direct numerical simulations of low-Reynolds number flows (see figure \ref{fig_model_coeff}).

\subsection{Zeroth-order conditional velocity gradient Laplacian}

At order zero in $\Rey_\gamma$, the equation governing the velocity gradient dynamics \eqref{eq_Langevin_A} is linear, and the resulting flow has Gaussian statistics.
For a multi-point Gaussian random field, the conditional Laplacian reduces to a linear damping \citep{Wilczek2014}. Therefore the conditional Laplacian in dimensional variables is of the form
\begin{align}
\left\langle\bar{\nabla}^2 \bar{\tens{A}}(\bar{\vect{x}},\bar{t})\middle|\bar{\tens{A}}\right\rangle = -\frac{\bar{\tens{A}}}{\bar{\gamma}_0^2} + \ordof\left(\Rey_\gamma\right).
\label{eq_V_Gauss}
\end{align}
and in the non-dimensional variables \eqref{def_nd_variables}, we just have
$\langle\nabla^2 \tens{A}|\tens{A}\rangle \sim -\tens{A}$.
The conditional velocity gradient Laplacian for a Gaussian random field contributes to the model coefficients \eqref{eq_tens_repr} at order zero in Reynolds number, while the higher-order viscous corrections require modelling.
The reference length $\bar{\gamma}_0$ depends on the spatial correlation of the forcing, and we determine it in Appendix \ref{app_Wyld}.

\subsection{Zeroth-order conditional anisotropic pressure Hessian}

For a Gaussian random field, that is at order zero in Reynolds number, the conditional traceless/anisotropic pressure Hessian takes a simple form \citep{Wilczek2014}
\begin{align}
\left\langle \widetilde{\tens{H}}(\vect{x},t)\middle|\tens{A}\right\rangle = -\frac{2}{7}\widetilde{\tens{SS}} - \frac{2}{5}\widetilde{\tens{WW}}
+
h_4 \left(\tens{SW} - \tens{WS}\right)
+\ordof(\Rey_\gamma),
\label{eq_H_Gauss}
\end{align}
while the local/isotropic part of the Hessian is specified by the incompressibility condition $\Tr(\tens{H})=-\Tr(\tens{A}^2)$.
The conditional anisotropic pressure Hessian for a Gaussian random field contributes to the model coefficients \eqref{eq_tens_repr} at order one in Reynolds number. Therefore, we know from \cite{Wilczek2014} the exact first-order correction to the gradient dynamics at small $\Rey_\gamma$ due to the pressure Hessian.

The coefficient $h_4$ weighing $\tens{B}^4 = \tens{SW} - \tens{WS}$ in \eqref{eq_H_Gauss} depends on the structure of the Gaussian flow through the correlation function, but previous works have shown that it does not contribute to the single-point statistics of the velocity gradient. This can be seen geometrically since $\tens{B}^4$ rotates the strain-rate eigenframe while leaving unchanged the vorticity orientation with respect to the eigenframe itself \citep{Carbone2020b}. It can also be seen from the Fokker-Planck equation for the gradient PDF \eqref{eq_FPE_inv} since $\tens{B}^4$ contracts to zero with all the other basis tensors \citep{Leppin2020}, as from the fourth row/column of the metric tensor \eqref{eq_def_met_tens}. Therefore, we can ignore the contributions from $\tens{B}^4$ as long as we are concerned with single-time/single-point statistics.

\subsection{Tensor representation of the conditional averages and resulting velocity gradient model}

We model the higher-order contributions from the unclosed terms by employing the general expression \eqref{eq_tens_repr}, truncated at degree two in the velocity gradient.
We consider the basis tensors \eqref{eq_def_BT} from $\tens{B}^1$ to $\tens{B}^6$, by keeping in mind that $\tens{B}^4$ can be ignored as long as we focus on single-point statistics. Also, we assume that the coefficients $\gamma_n$ in \eqref{eq_tens_repr} are constant. These hypotheses, together with the exact expression of the zeroth-order conditional averages (\ref{eq_V_Gauss},\ref{eq_H_Gauss}), yield the following representation of the drift term \eqref{eq_tens_repr}
\begin{align}
&\Rey_\gamma\left(\tens{A}\tens{A} + \left\langle \tens{H}(\vect{x},t)\middle|\tens{A}\right\rangle  \right)
-\left\langle\nabla^2 \tens{A}(\vect{x},t)\middle|\tens{A}\right\rangle =
\tens{A} - 
\Rey_\gamma^2\left[\delta_1\tens{S}  +
\delta_2\tens{W}\right]  +\nonumber\\
+& \Rey_\gamma\left[\left(\frac{5}{7}-\delta_3\right)\widetilde{\tens{SS}} +
\gamma_4\left(\tens{SW}-\tens{WS}\right) +
\left(1-\delta_5\right)\left(\tens{SW}+\tens{WS}\right) +
\left(\frac{3}{5}-\delta_6\right)\widetilde{\tens{WW}}\right]
\label{eq_higher_order}
\end{align}
where the constant coefficients $\delta_i$ are of order one and are to be determined. We include second-order terms in $\Rey_\gamma$ to keep the variance of the gradients constant for all Reynolds numbers, as is the case for the stochastically driven Navier-Stokes equations \eqref{eq_NS_ndim}. The powers in the Reynolds numbers in equation \eqref{eq_higher_order} have been chosen so that the model equations remain unchanged under the transformation
$\bar{t}\to -\bar{t}$ and $\bar{\nu}\to-\bar{\nu}$, 
as is the case for the Navier-Stokes equations \eqref{eq_NS_ndim}.
We will test  a-posteriori the trend of the model coefficients in terms of the Reynolds number against the DNS data (see figure \ref{fig_model_coeff}).

The constant model parameters $\delta_1,\delta_2,\delta_3,\delta_5,\delta_6$ remain to be determined. To do this, we use the two Betchov homogeneity constraints \citep{Betchov1956}, the perturbation theory at small Reynolds for the full Navier-Stokes equations \citep{Wyld1961}, together with the constant dissipation rate imposed by the stochastic forcing \citep{Novikov1965}.
This results in constraints on the average of the velocity gradient invariants
\begin{subequations}
\begin{align}
\avg{\inv_1} &= \frac{1}{2}, &&
\avg{\inv_1+\inv_2} = 0, &&
\avg{\inv_3+3\inv_4} = 0,\label{avg_constr_Betchov}
\\
\avg{\inv_3} &= S_3\Rey_\gamma, &&
\avg{\inv_5} = -\frac{1}{12} + X_5\Rey_\gamma^2
\label{avg_constr_Wyld}
\end{align}
\label{eq_avg_constr}
\end{subequations}
where $S_3$ and $X_5$ are constant parameters.
The first relation in \eqref{avg_constr_Betchov} follows from the constant variance of the gradients imposed by the stochastic forcing. It implies that the Kolmogorov time scale $\tau_\eta=1/\sqrt{2\avg{\inv_1}}$ is unitary in the non-dimensional variables \eqref{def_nd_variables}.
The other relations in \eqref{avg_constr_Betchov} are the two independent Betchov homogeneity constraints that can be formulated using solely the velocity gradient \citep{Carbone2022}.
The homogeneity relations \eqref{avg_constr_Betchov} have been already employed to reduce the number of parameters in numerical simulations of velocity gradient models at high Reynolds numbers \citep{Leppin2020}, and here we can impose those constraints analytically, as we will see below.
Relations \eqref{avg_constr_Wyld} follow from a low-Reynolds-number expansion of the Navier-Stokes equations \citep{Wyld1961}. The quantities $S_3$ and $X_5$ represent, respectively, the rate of change of the third- and fourth-order moments of the velocity gradient with the Reynolds number, starting from a Gaussian zeroth-order configuration. These coefficients depend on the forcing correlation, and we determine them in Appendix \ref{app_Wyld}.

In the constraints \eqref{eq_avg_constr} the brackets indicate the ensemble average, that is, for a generic function of the velocity gradient $\varphi(\tens{A})$,
\begin{align}
\avg{\varphi(\tens{A})} = \int\de\tens{A}f\left(\inv(\tens{A});\delta_i\right)\varphi(\tens{A}),
\label{eq_ens_avg_A}
\end{align}
where $f(\inv;\delta_i)$ is the PDF of the velocity gradient governed by the FPE \eqref{eq_FPE_inv}, with the drift term \eqref{eq_tens_repr}.
The FPE \eqref{eq_FPE_inv} admits perturbative exact solutions at small Reynolds numbers, as described in further detail in the next section. This analytic solution $f(\inv;\delta_i)$ depends parametrically on the model coefficients $\delta_i$, and it allows us to compute the ensemble averages in \eqref{eq_avg_constr} analytically.
Therefore, the five constraints \eqref{eq_avg_constr} constitute a linear system of five equations for the five model parameters $\delta_1,\delta_2,\delta_3,\delta_5,\delta_6$. The solution of this linear system is
\begin{align}
\delta_1 &= - \frac{1824 S_{3}^{2}}{35} - \frac{144 X_{5} }{7}&&
\delta_2 = \frac{96 S_{3}^{2} }{7} + \frac{240 X_{5}}{7} \nonumber\\
\delta_3 &= \frac{5}{7} + \frac{120 S_{3} }{7} &&
\delta_4 = \gamma_4 -h_4  \nonumber\\
\delta_5 &= 1 - \zeta_5 - \frac{72 S_{3} }{7} - \frac{180 X_{5} }{7 S_{3}} &&
\delta_6 = \frac{3}{5} - \frac{6 \zeta_5}{5} - \frac{936 S_{3} }{35}  - \frac{216 X_{5} }{7 S_{3}}.
\label{eq_model_coeff}
\end{align}
Additionally, the noise variance $\sigma^2=1/15$ is fixed by the unitary Kolmogorov time constraint at $\Rey_\gamma=0$ (at which the FPE \eqref{eq_FPE_inv} with the drift \eqref{eq_higher_order} is an Ornstein–Uhlenbeck process). We assume that the noise variance $\sigma^2$ is independent of the Reynolds number.

We have now determined all the single-time/single-point model coefficients, $\delta_i$ and $\sigma$. The model features the exact first-order contribution in $\Rey_\gamma$ from the pressure Hessian, while the modelling hypotheses concern the representation of the higher-order corrections. The resulting model for the velocity gradient single-time statistics does not feature any free parameter, not requiring any parameter scan to match the DNS results.
Still, there are two free gauge parameters, namely $\gamma_4$ and $\zeta_5$, which do not affect single-time statistics, but only multi-time correlations.
This gauge stems from the fact that the Gaussian and  isotropic PDF with unitary Kolmogorov time scale \citep[e.g.][]{Wilczek2014}
\begin{align}
    f_0 = \frac{225 \sqrt{5}}{\pi^{4}} \exp\left[- \left( 4\delta_{ip}\delta_{jq}+ \delta_{iq}\delta_{jp}\right)A_{ij}A_{pq}\right] 
\end{align}
solves the non-linear steady-state FPE
\begin{align}
\frac{\partial}{\partial A_{ij}} \Bigg[ \left(A_{ij} + \Rey_\gamma \zeta_5\left( B^5_{ij}+\frac{6}{5}B^6_{ij}\right) + \Rey_\gamma \gamma_4 B^4_{ij}\right)f_0 + \frac{1}{30}\left(4\frac{\partial f_0}{\partial A_{ij}}  - \frac{\partial f_0}{\partial A_{ji}} \right)\Bigg] = 0,
\end{align}
for all $\gamma_4$ and $\zeta_5$.
A particular case of this gauge for the zeroth-order Gaussian solution has been observed by \cite{Leppin2020}.
We will determine the gauge parameters $\gamma_4$  and $\zeta_5$ in section \ref{sec_trajectories}, where we focus on multi-time statistics, with the aid of DNS data.

\section{Analytic approximation of the velocity gradient PDF at small Reynolds numbers}
\label{sec_solution}

The Fokker-Planck equation \eqref{eq_FPE_inv} with the drift term specified by \eqref{eq_higher_order} admits perturbative solutions in the Reynolds number. The solution is expanded up to second order
\begin{align}
f(\inv) \sim f_0(\inv) + \Rey_\gamma f_1(\inv) + \Rey_\gamma^2 f_2(\inv),
\label{eq_f_exp_Re}
\end{align}
and we solve for $f_i$ at all orders, in the form of a polynomial of the invariants times the zeroth-order Gaussian solution. In particular, plugging the expansion \eqref{eq_f_exp_Re} into the FPE \eqref{eq_FPE_inv}, comparing terms of the same order in Reynolds number, and imposing the average constraints \eqref{eq_avg_constr}
yields the following asymptotic solution of the FPE at small $\Rey_\gamma$
\begin{subequations}
\begin{align}
f_0 &= \frac{225 \sqrt{5}}{\pi^{4}} e^{- 5 \inv_1 + 3 \inv_2}
\label{eq_FPE_sol_0}\\
f_1 &= \frac{3600 \sqrt{5} S_{3} \left(25 \inv_3 - 21 \inv_4\right)}{7 \pi^{4}} e^{- 5 \inv_1 + 3 \inv_2}
\label{eq_FPE_sol_1}\\
f_2 &= \frac{720 \sqrt{5}}{49 \pi^{4}} \big(- 16320 S_{3}^{2} \inv_1 \inv_2 - 6860 S_{3}^{2} \inv_1 - 1344 S_{3}^{2} \inv_2^{2} - 140 S_{3}^{2} \inv_2 + 50000 S_{3}^{2} \inv_3^{2}  + \nonumber\\
&- 84000 S_{3}^{2} \inv_3 \inv_4 + 35280 S_{3}^{2} \inv_4^{2} + 42240 S_{3}^{2} \inv_5 + 2240 S_{3}^{2} - 22950 X_{5} \inv_1 \inv_2 - 1575 X_{5} \inv_1 + \nonumber\\ &-1890 X_{5} \inv_2^{2} - 1575 X_{5} \inv_2 + 59400 X_{5} \inv_5\big) e^{- 5 \inv_1 + 3 \inv_2}.
\label{eq_FPE_sol_2}
\end{align}
\label{eq_FPE_sol}
\end{subequations}
We remark that while the solution \eqref{eq_FPE_sol} is only asymptotic in Reynolds number, the terms $f_i$ in \eqref{eq_FPE_sol} solve exactly, for all $\inv_k$, the expanded partial differential equation \eqref{eq_FPE_inv} at each order in the Reynolds number. Therefore, while we deal with an asymptotic expansion at small Reynolds number, there is no explicit assumption on the magnitude of the velocity gradients.
To assess the asymptotic solution \eqref{eq_FPE_sol_1} by symbolic computation, we rewrote the terms in \eqref{eq_FPE_sol_2} as functions of the Cartesian components of the velocity gradient, inserted the resulting expression into the Cartesian FPE \eqref{eq_FPE_Cart}, and checked that the remainder is zero up to second order in Reynolds number.

The main advantage of the simple expansion \eqref{eq_FPE_sol} is that it gives full analytic access to the relevant moments of the PDF. This allows us to investigate analytically the onset of non-Gaussianity at small Reynolds number.
Instead, a shortcoming of this expansion is that the solution is not positive for all $\inv_k$, as it should be for a proper PDF. However, this positivity breaking happens when the polynomial pre-factors in \eqref{eq_f_exp_Re} vanish, that is, when the gradients are of the order of $\Rey_\gamma^{-1/3}$. At such large gradients, the Gaussian factor $\exp(- 5 \inv_1 + 3 \inv_2)$ has already strongly decayed and the consequent error is very small. The comparison with the numerical results in Section \ref{sec_results} will confirm that this positivity issue is indeed negligible at low Reynolds numbers.

An alternative way to obtain an asymptotic solution of the FPE \eqref{eq_FPE_inv} consists of the effective action method, proposed in \cite{Martin1973} for stochastic differential equations. This method yields solutions of the form $\exp(-\mathcal{S}(\inv))$ where $\mathcal{S}$ is the effective action, featuring higher-order polynomials in $\inv_k$ and $\Rey_\gamma$. Such a PDF does not give analytic access to the moments, which are usually computed numerically via Monte-Carlo sampling \citep{Moriconi2014}.
The effective action involves renormalized noise variance and model coefficients \citep{Apolinario2019} aiming to improve the accuracy and range of validity of the analytic predictions. Our model targets $\Rey_\gamma$ very small, and the equations themselves are only valid for low Reynolds numbers.
Therefore, in this setup, the simple expansion yielding the solution \eqref{eq_FPE_sol} seems appropriate, and we leave more advanced approaches for future work.

\subsection{Volume elements and the moments of the velocity gradient}

The solution \eqref{eq_FPE_sol} allows us to compute ensemble averages analytically. The ensemble average \eqref{eq_ens_avg_A} is conveniently computed in the strain-rate eigenframe. We express all the invariants $\inv_k$ in terms of the strain-rate eigenvalues $\lambda_i$ and vorticity principal components $\omega_i=\vect{v}_i\bcdot\vect{\omega}$, where $\vect{v}_i$ are the strain-rate eigenvectors. Additionally, we use coordinates in the strain-rate eigenframe based on the vorticity magnitude  $\omega=\|\vect{\omega}\|$ and the alignments between the vorticity vector and strain-rate eigenvectors,  $\hat{\omega}_i\equiv \omega_i/\omega$.
This procedure transforms the integral \eqref{eq_ens_avg_A} into
\begin{align}
\avg{\varphi(\inv)} =
\int\de\tens{S}\int\de\omega_1\de\omega_2\de\omega_3 f(\inv) \varphi(\inv)
\end{align}
with the integration operators in the strain-rate eigenframe specified by
\begin{subequations}
\begin{align}
\int\de\tens{S}  &= 
\int_{0}^{\infty}\de\lambda_2\int_{\lambda_2}^{+\infty}\de\lambda_1 J_S(\lambda)
+
\int_{-\infty}^{0}\de\lambda_2\int_{-2\lambda_2}^{+\infty}\de\lambda_1 J_S(\lambda)
\label{eq_ens_avg_str}
\\
\int\de\omega_1\de\omega_2\de\omega_3 &= 
\int_{0}^{+\infty}\de\omega^2\int_0^1\de\hat{\omega}_1^2\int_0^1\de\hat{\omega}_2^2 J_\omega(\omega,\hat{\omega}), \label{eq_ens_avg_omg}
\end{align}
\label{eq_ens_avg}
\end{subequations}
and where the invariants are also expressed as functions of the strain-rate variables,
\begin{align}
\inv_1 &= \sum_i \lambda_i^2, &&
\inv_2 = -\frac{\omega^2}{2}, \nonumber\\
\inv_3 &= \sum_i \lambda_i^3, &&
\inv_4 = \frac{\omega^2}{4}\sum_i \lambda_i\hat{\omega}_i^2, &&
\inv_5 = \frac{\omega^2}{4}\sum_i \lambda_i^2\left(\hat{\omega}_i^2-1\right).
\label{eq_inv_eigS}
\end{align}
Due to the statistical isotropy, rotations of the strain-rate eigenframe can be integrated out. This reduces the dimensionality of the ensemble average integral from eight (i.e., the independent components of the traceless $\tens{A}$ in \eqref{eq_ens_avg_A}) to five (i.e., the independent strain-rate eigenvalues and the vorticity principal components in \eqref{eq_ens_avg}). However, integrating out rotations also comes at the cost of introducing volume elements \citep{Livan2018}, namely $J_S$ and $J_\omega$ in equation \eqref{eq_ens_avg_str}.
The volume element $J_S$ associated with the transformation from standard Cartesian reference frame to the strain-rate eigenframe consists of the Wigner repulsion term \citep{Wigner1955}
\begin{align}
J_S = 2\pi^2\left|(\lambda_1-\lambda_2)(\lambda_2-\lambda_3)(\lambda_1-\lambda_3)\right| = \sqrt{2}\pi^2\sqrt{\inv_1^3-6\inv_3^2}.
\label{eq_vol_el_S}
\end{align}
The absolute value in \eqref{eq_vol_el_S} drops when employing the ordered strain-rate eigenvalues $\lambda_1>\lambda_2>\lambda_3$ as integration variables. Due to incompressibility, $\lambda_1+\lambda_2+\lambda_3=0$, and we have two possible configurations, $\lambda_1>\lambda_2>0$ or $\lambda_1>-2\lambda_2>0$, which specify the integration bounds in \eqref{eq_ens_avg_str}.
Finally, going from Cartesian coordinates to the vorticity magnitude/orientation in the strain-rate eigenframe introduces the volume element
\begin{align}
J_\omega =
\frac{1}{8}\sqrt{\frac{\omega^2}{\hat{\omega}^2_1\hat{\omega}^2_2\hat{\omega}^2_3}}
\label{eq_vol_el_omg}
\end{align}
where $\hat{\omega}_3^2 = 1-\hat{\omega}_1^2-\hat{\omega}_2^2$, and we can consider only positive values of the vorticity principal components $\hat{\omega}_i$ since the invariants $\inv_k$ \eqref{eq_inv_eigS} are even functions of the vorticity.
\section{Comparison of single-point/single-time statistics}
\label{sec_results}

We compare the single-time/single-point velocity gradient statistics resulting from our model and from direct numerical simulations at low Reynolds numbers.
We point out several qualitative changes in the dynamics and statistical geometry of the gradient as the Reynolds number increases till a transition to turbulence.
The DNS setup is detailed in Appendix \ref{app_DNS_details}, while all the theoretical predictions follow by integrating out variables from the asymptotic solution \eqref{eq_f_exp_Re}, making use of the expressions for the ensemble average \eqref{eq_ens_avg}.
The DNS and model parameters 
are in one-to-one correspondence. In particular, the zeroth-order conditional viscous damping and the correlations of the forcing are the same in the model and in the DNS, as described in the Appendices \ref{app_DNS_details} and \ref{app_Wyld}.

\subsection{Moments of the velocity gradient invariants}

Increasing the Reynolds number starting from zero leads to the onset of the skewness, intermittency and preferential alignments in the gradient statistics. We analyze those three features separately by looking at the strain-rate and vorticity statistics from the strain-rate eigenframe viewpoint \citep{Dresselhaus1992,Tom2021}.

\begin{figure}
  \centerline{
\begin{overpic}[width=\textwidth]{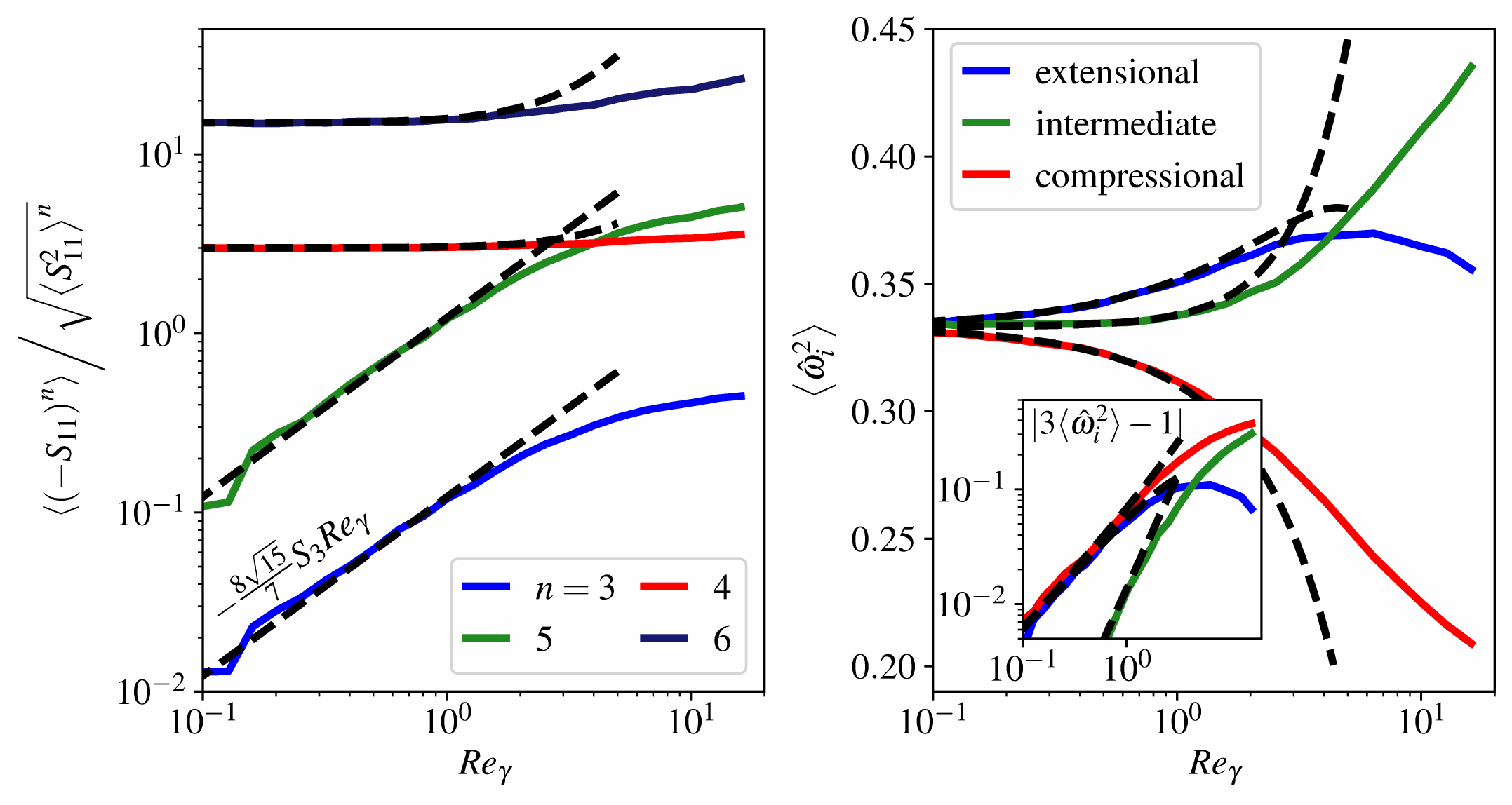}
\put(30,52){(\textit{a})}
\put(80,52){(\textit{b})}
\end{overpic}
}
  \caption{Onset of skewness, intermittency and alignments at low Reynolds number, in terms of the expansion parameter $\Rey_\gamma$. (\textit{a}) Normalized moments of the strain-rate longitudinal component $S_{11}$. Note that we plot $\langle (-S_{11})^n\rangle$ since the odd-order moments are negative.
(\textit{b}) Alignments between the vorticity and the strain-rate eigenvectors.
Solid coloured lines are from the DNS, while black dotted lines are the analytic predictions from our low-Reynolds number model.}
\label{fig_mom_invs}
\end{figure}

Figure \ref{fig_mom_invs}(a) shows the normalized moments of the strain-rate longitudinal components as a function of the Reynolds number $\Rey_\gamma$.
The moments of the invariants are related to the moment of the longitudinal strain-rate component through \citep{Betchov1956,Davidson2015}
\begin{align}
    \avg{S_{11}^2} = \frac{2}{15}\avg{\inv_1}, &&
    \avg{S_{11}^3} = \frac{8}{105}\avg{\inv_3}, &&
    \avg{S_{11}^4} = \frac{4}{105}\avg{\inv_1^2}.
\end{align}
The skewness of the strain rate is quantified via the third-order moment of the strain-rate $\avg{\inv_3}$, while the even and higher-order moments, e.g.~$\avg{\inv_1^2}$, quantify the intermittency. 
After the rapid initial increase, the skewness of the strain-rate longitudinal component saturates towards its typical high-Reynolds-number value of $-0.5$/$-0.6$. At large Reynolds number, this skewness is predicted to be constant by a renormalization approach \citep{Yakhot1986}, while it weakly increases with the Reynolds number with an exponent of order $0.1$ in numerical simulations \citep{Ishihara2009}.
At $\Rey_\gamma=\ordof(1)$, the strain-rate statistics already display a remarkable skewness, while the intermittency is negligible.
This is because the third-order moments, e.g.~$\avg{\inv_3}$, grow linearly with the Reynolds number, while even-order moments, e.g.~$\avg{\inv_1^2}$, grow quadratically.
Recent works \citep{Yakhot2017,Gotoh2022}, have identified a transition to turbulence based on the sudden growth of the even velocity gradient moments and the onset of a power law as their growth begins.
The DNS data indicate that the transition 
to turbulent scalings starts at $\Rey_\lambda\approx 9$ \citep{Yakhot2017}, corresponding to $\Rey_\gamma\approx 5$ (see figure \ref{fig_DNS_setup}).
Our velocity gradients model encodes by construction the different scaling of odd and even moments at low Reynolds number. Indeed, those scalings can be analytically derived through a small-Reynolds-number expansion of the Navier-Stokes equations, as outlined in Appendix \ref{app_Wyld}. 

Figure \ref{fig_mom_invs}(b) shows a striking qualitative change in the statistical geometry of the velocity gradient at low Reynolds numbers: the switching in the preferential alignments between the strain rate and the vorticity.
The alignments are characterized through the normalized vorticity components in the strain-rate eigenframe $\hat{\omega}_i=\vect{v}_i\bcdot\vect{\omega}/\|\vect{\omega}\|$ (where $\vect{v}_i$ are the strain-rate eigenvectors), ordered based on the corresponding strain-rate eigenvalue $\lambda_i$, with $\lambda_1\ge\lambda_2\ge\lambda_3$.
The vorticity aligns with the most extensional strain-rate eigenvector at low Reynolds number, and only at $\Rey_\gamma\gtrapprox 5$ it aligns with the intermediate eigenvector.
At $\Rey_\gamma<5$, the alignments are as if the vorticity was a material line in the fluid flow subject to a persistent strain.
Then, at $\Rey_\gamma\approx 5$, the transition to turbulence begins: the somewhat counter-intuitive alignment between the intermediate strain-rate eigenvector and the vorticity establishes, as observed in fully developed turbulence \citep{Ashurst1987}. 
The onset of this peculiar alignment at a higher Reynolds number is consistent with the local nonlinearities in the velocity gradient dynamics becoming more relevant. For example, the simple restricted Euler model \citep{Vieillefosse1982}, in which the gradient is driven only by the local/anisotropic part of the nonlinear term, also displays alignment between the vorticity and the intermediate strain-rate eigenvector while approaching a finite-time singularity.

By integrating out the strain-rate eigenvalues and vorticity magnitude from the asymptotic solution of the FPE \eqref{eq_f_exp_Re}, with the ensemble average expressed in the form of \eqref{eq_ens_avg}, we can get an analytic approximation of the vorticity principal components as a function of the Reynolds number
\begin{align}
\avg{\hat{\omega}_1^2} &= \frac{1}{3} - \frac{6 \sqrt{30} S_{3} \Rey_\gamma}{25 \sqrt{\pi}} + \frac{4224 S_{3}^{2} \Rey_\gamma^{2}}{1225} + \frac{1188 X_{5} \Rey_\gamma^{2}}{245}\nonumber\\
\avg{\hat{\omega}_2^2} &= \frac{1}{3} - \frac{8448 S_{3}^{2} \Rey_\gamma^{2}}{1225} - \frac{2376 X_{5} \Rey_\gamma^{2}}{245}\nonumber\\
\avg{\hat{\omega}_3^2} &= \frac{1}{3} + \frac{6 \sqrt{30} S_{3} \Rey_\gamma}{25 \sqrt{\pi}} + \frac{4224 S_{3}^{2} \Rey_\gamma^{2}}{1225}  + \frac{1188 X_{5} \Rey_\gamma^{2}}{245}.
\label{eq_sol_omg_align}
\end{align}
Both $\avg{\hat{\omega}_1^2}$ and $\avg{\hat{\omega}_2^2}$ start from a random uniform value $\avg{\hat{\omega}_i^2}=1/3$ at zero Reynolds number. Then $\avg{\hat{\omega}_1^2}$ grows linearly with $\Rey_\gamma$ while $\avg{\hat{\omega}_2^2}$ grows only quadratically.
The preferential alignments are closely related to the onset of the strain-rate skewness. Indeed, the fact that $S_3<0$ (analytically derived through the Wyld expansion in Appendix \ref{app_Wyld}) not only implies an average direct energy cascade, but it also implies that the vorticity preferentially aligns with the most extensional direction at low Reynolds number, as seen from equation \eqref{eq_sol_omg_align}.
However, $\avg{\hat{\omega}_1^2}$ has a strong negative second-order contribution in Reynolds and eventually, $\avg{\hat{\omega}_2^2}$, which has a positive growth rate, takes over.
The analytic approximation \eqref{eq_sol_omg_align} suggests that the growth of $\avg{\hat{\omega}_2^2}$ with $\Rey_\gamma$ can only take place if $X_5$ is negative,
and its magnitude is large enough compared to $S_3^2$.
Furthermore, it is known that the vorticity avoids the compressional direction in turbulence \citep{Ashurst1987}.
This lack of alignment establishes already at very low Reynolds numbers, and it strongly depends on the fact that $S_3$ and $X_5$ are both negative. Our model captures this lack of alignment since $\avg{\hat{\omega}_3^2}$ in \eqref{eq_sol_omg_align} has negative first- and second-order contributions, so it quickly decreases with the Reynolds number.

As the comparison between the DNS and model predictions for the vorticity--strain rate alignments shows, our low-Reynolds number model is quantitatively accurate up to $\Rey_\gamma\approx 1$, in qualitative agreement with the DNS until $\Rey_\gamma\approx 5$ and then breaks down, as expected for a weak-coupling perturbative approach.
The results highlight many qualitative changes as the Reynolds number increases, consistent with the observations in recent works \citep{Yakhot2017,Gotoh2022}. However, the Taylor Reynolds number is usually employed to investigate the transition to turbulence in the literature, while we localize the transition at $\Rey_\gamma\approx 5$. In our small-Reynolds number analysis, the actual expansion parameter is $\Rey_\gamma$, which is non-trivially related to the Taylor Reynolds number, their relation depending on the spatial correlations of the forcing.

\subsection{Strain-rate statistics}

In order to better understand the origin of the strain-rate skewness and to get insight into the preferential configuration of the strain-rate eigenvalues observed in fully developed turbulence \citep{Betchov1956}, we analyze the PDF of the strain-rate tensor using elements of the theory of random matrices \citep{Livan2018}.
At infinitesimally small Reynolds number, the velocity field is multi-point Gaussian, and the strain rate is an orthogonal Gaussian random matrix.
Gaussian random matrices are well-established tools in many research areas \citep{Wigner1955,Dyson1962}, but less frequently employed in fluid dynamics. For example, the distribution of the dissipation rate corresponding to a Gaussian strain rate has been only recently re-derived in this field \citep{Gotoh2022}.

\begin{figure}
  \centerline{
\begin{overpic}[width=.9\textwidth]{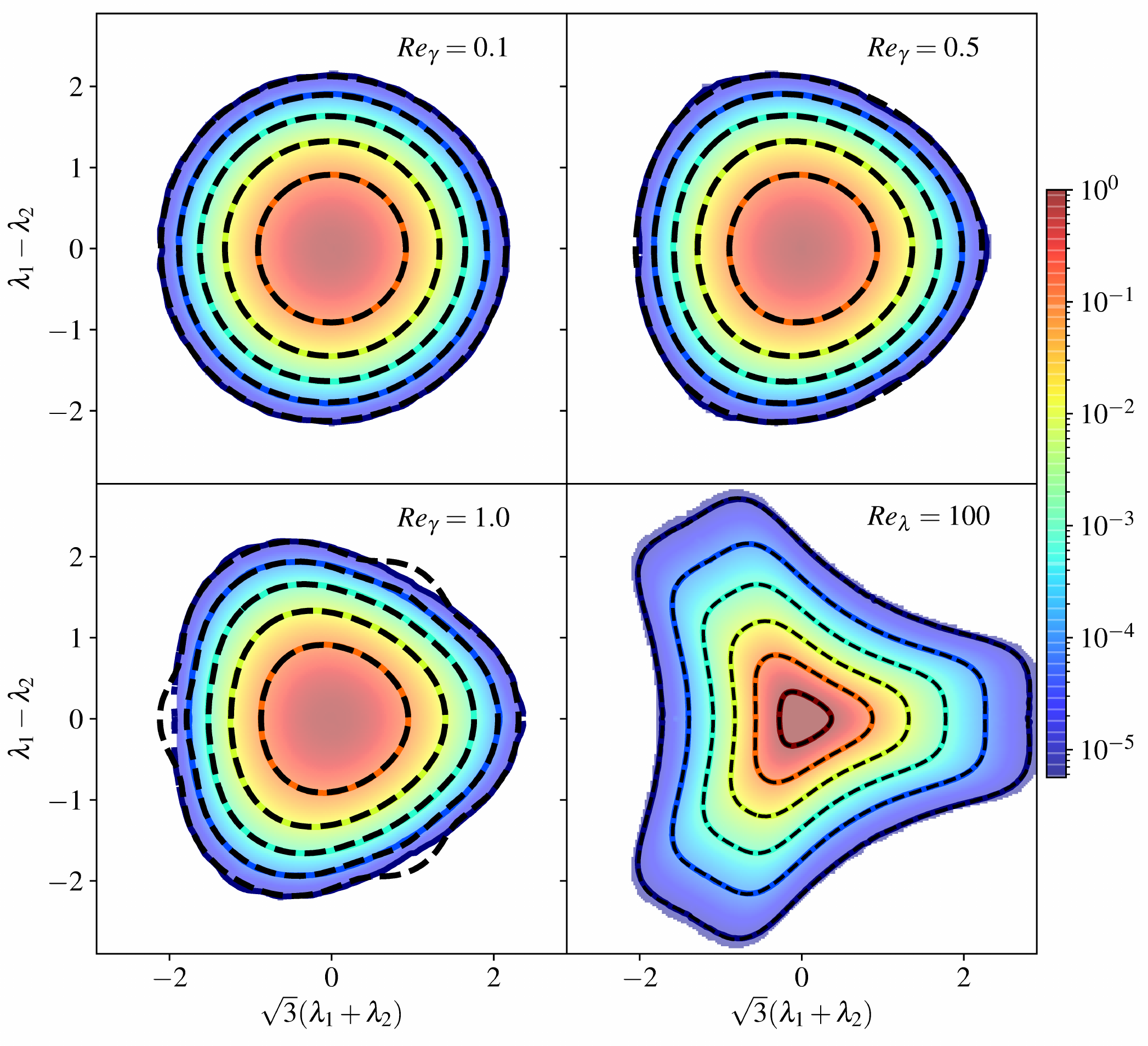}
\end{overpic}
  }
  \caption{PDF of the strain rate $f_S(\lambda)$ parameterized through its unordered and rescaled eigenvalues $\lambda_i$, at various Reynolds numbers. The colormap and coloured solid contours are from the DNS, while the dotted black lines indicate the corresponding analytic prediction from our low-Reynolds number model.
	The thin dotted lines in the high-Reynolds number plot indicate the empirical parameterization \eqref{eq_str_pdf_cont}.
	The colormap is $\log_{10}$ scale, and the contours are equispaced in $\log_{10}$ scale with unitary increments.}
\label{fig_PDF_lam_str}
\end{figure}

Due to incompressiblity and statistical isotropy, the strain-rate PDF is parameterized through two quantities, namely the two independent (unordered) strain-rate eigenvalues $\lambda_{1}$ and $\lambda_{2}$, or the two principal invariants
$\inv_1 = 2(\lambda_1^2+\lambda_2^2+\lambda_1\lambda_2)$ and $\inv_3 = 3\lambda_1\lambda_2(\lambda_1+\lambda_2)$. 
Therefore, changing coordinates from the standard Cartesian basis to the strain-rate eigenframe brings a remarkable dimensionality reduction. This change of coordinates implies that the PDF of the strain-rate PDF $f_S(\lambda(\tens{S}))$ is related to the PDF of its eigenvalues $\rho_{\lambda} (\lambda)$ through \citep{Livan2018}
\begin{align}
    f_S(\lambda) = \frac{\rho_\lambda(\lambda)}{J_S(\lambda)},
\label{eq_def_f_S_weigh}
\end{align}
where  $J_S$ is the volume element of the Stiefel manifold \citep{James1977}, defined in \eqref{eq_vol_el_S}.
Since the volume element $J_S$ has a convoluted form, the PDF of the strain rate parameterized through the eigenvalues, $f_S(\lambda)$, is visually much simpler than the PDF of the eigenvalues themselves $\rho(\lambda)$. We will therefore plot and discuss $f_S(\lambda)$, shown in figure \ref{fig_PDF_lam_str} for various Reynolds numbers.
Furthermore, we 
employ rescaled sum and difference of the eigenvalues, so that the contours of the strain-rate PDF at vanishingly small Reynolds number are circles.
The contours of the PDF in figure \ref{fig_PDF_lam_str} are logarithmically equispaced, the coloured solid lines are from the DNS, and the dotted black lines represent the analytic prediction obtained by integrating out the vorticity from \eqref{eq_f_exp_Re} in the fashion of \eqref{eq_ens_avg}
\begin{align}
&f_S(\lambda) =
\Bigg(
\frac{50 \sqrt{30}}{\pi^{\frac{5}{2}}}
-\Rey_\gamma \frac{60000 \sqrt{30} S_{3}}{7 \pi^{\frac{5}{2}}} \lambda_{1} \lambda_{2} \left(\lambda_{1} + \lambda_{2}\right) + \Rey_\gamma^2 \frac{20000 \sqrt{30} S_{3}^{2}}{49 \pi^{\frac{5}{2}}} \bigg(1800 \lambda_{1}^{4} \lambda_{2}^{2} + \nonumber\\
&+
3600 \lambda_{1}^{3} \lambda_{2}^{3} + 1800 \lambda_{1}^{2} \lambda_{2}^{4} - 42 \lambda_{1}^{2} - 42 \lambda_{1} \lambda_{2} - 42 \lambda_{2}^{2} + 7\bigg) 
\Bigg) e^{- 10 \lambda_{1}^{2} - 10 \lambda_{1} \lambda_{2} - 10 \lambda_{2}^{2}}.
\label{eq_sol_str_pdf_lam}
\end{align}
As the Reynolds number increases, the contours of the strain-rate PDF in figure \ref{fig_PDF_lam_str} transition from a circular to a triangular shape. The contours elongate towards large and positive $\lambda_1+\lambda_2$, that is large and negative $\lambda_3$, while shrinking along $\lambda_1-\lambda_2$, especially when $\lambda_3$ is large. This shape indicates a preferential state with a negative and  
large
eigenvalue, with the other two being smaller, positive and close to each other. This configuration has been observed by \cite{Betchov1956} from a phenomenological analysis of the strain-rate dynamics in high-Reynolds number turbulence, and equation \eqref{eq_sol_str_pdf_lam} analytically shows the onset of this preferential configuration.
The skewness in the PDF \eqref{eq_sol_str_pdf_lam} is due to the first-order contribution in $\Rey_\gamma$, which amplifies the probability of regions in which the product $S_3\lambda_1\lambda_2\lambda_3$ is negative. This corresponds to a preferentially positive intermediate strain-rate eigenvalue $\lambda_2$ \citep{Lund1992}.

The analytic prediction \eqref{eq_sol_str_pdf_lam} captures the strain-rate PDF up to $\Rey_\gamma=\ordof(1)$. Discrepancies appear in the zero-measure regions in which two of the strain-rate eigenvalues coincide. This could be due to the choice of the basis tensors \eqref{eq_def_BT}, which are not linearly independent when the strain rate is in a degenerate configuration.

Figure \ref{fig_PDF_lam_str} also shows that the shape of the strain-rate PDFs at low and moderate Reynolds numbers qualitatively differ. The edges displayed by the strain-rate PDF at moderately large Reynolds numbers are not present at lower Reynolds numbers and constitute a high-Reynolds feature. The symmetry of the PDF with respect to the axes $\lambda_1=\lambda_2$, $\lambda_1=-\lambda_2/2$ and $\lambda_2=-\lambda_1/2$ remains at all Reynolds numbers, simply because the eigenvalues can be arbitrarily interchanged. Even at large Reynolds numbers however, the contours of the strain-rate PDF look surprisingly simple. In fact, they can be parameterized through a combination of the principal invariants, namely 
\begin{align}
	\alpha_1\left(f_S\right)\inv_1^3 + \alpha_2\left(f_S\right)\inv_3\inv_1^{3/2}+
	\alpha_3\left(f_S\right)\inv_3^2=1,
	\label{eq_str_pdf_cont}
\end{align}
where the quantities $\alpha_i$ depend on the contour level $f_S$ itself. This parameterization is shown in figure \ref{fig_PDF_lam_str} for the large-Reynolds number strain-rate PDF.

\begin{figure}
  \centerline{
  \includegraphics[width=\textwidth]{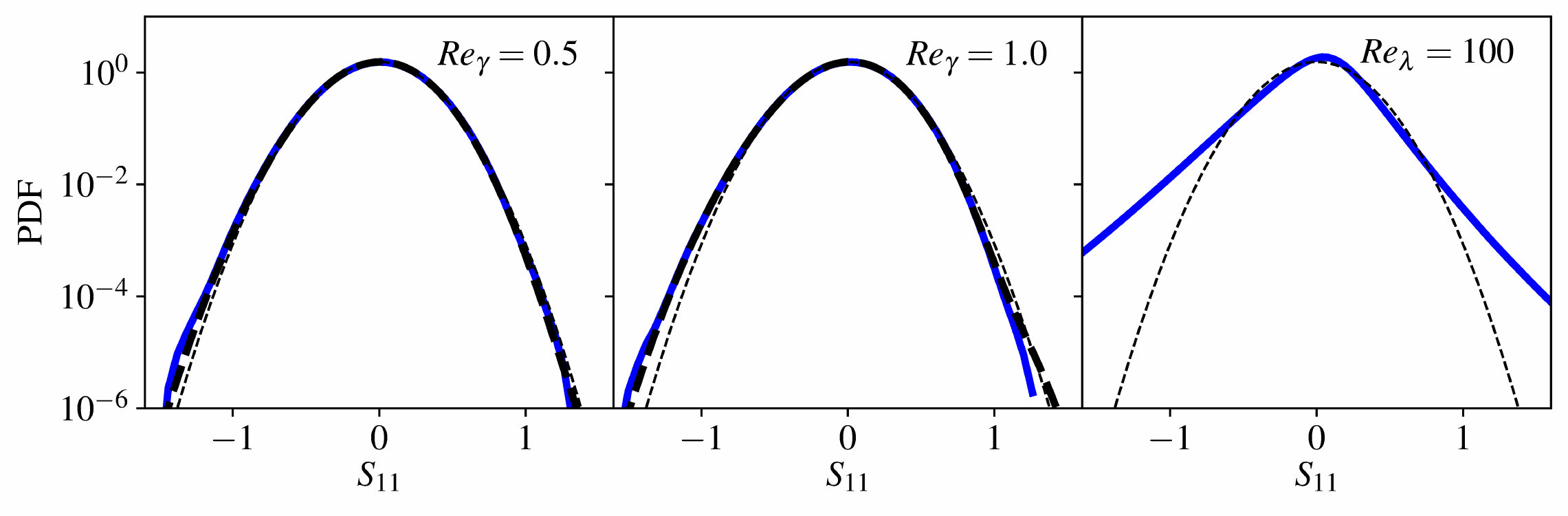}
  }
  \caption{PDF of the strain-rate longitudinal component at various Reynolds numbers. The solid lines are from the DNS, while the dotted black lines indicate the analytic prediction from our low-Reynolds number model. The thinner dotted line indicates the zeroth-order Gaussian PDF.}
\label{fig_pdf_a11}
\end{figure}

The onset of skewed and elongated tails of the strain-rate PDF results in the non-Gaussianity of the strain-rate longitudinal component PDF, shown in figure \ref{fig_pdf_a11}. As the Reynolds number increases, a left tail develops, while the right branch becomes slightly sub-Gaussian near the PDF core. Our model captures the slight increase of the left tail up to $\Rey_\gamma=\ordof(1)$, even though the non-Gaussianity at such low Reynolds numbers is very mild compared to the non-Gaussianity in high-Reynolds number turbulence.

\subsection{Vorticity statistics}

\begin{figure}
  \centerline{
  \includegraphics[width=.9\textwidth]{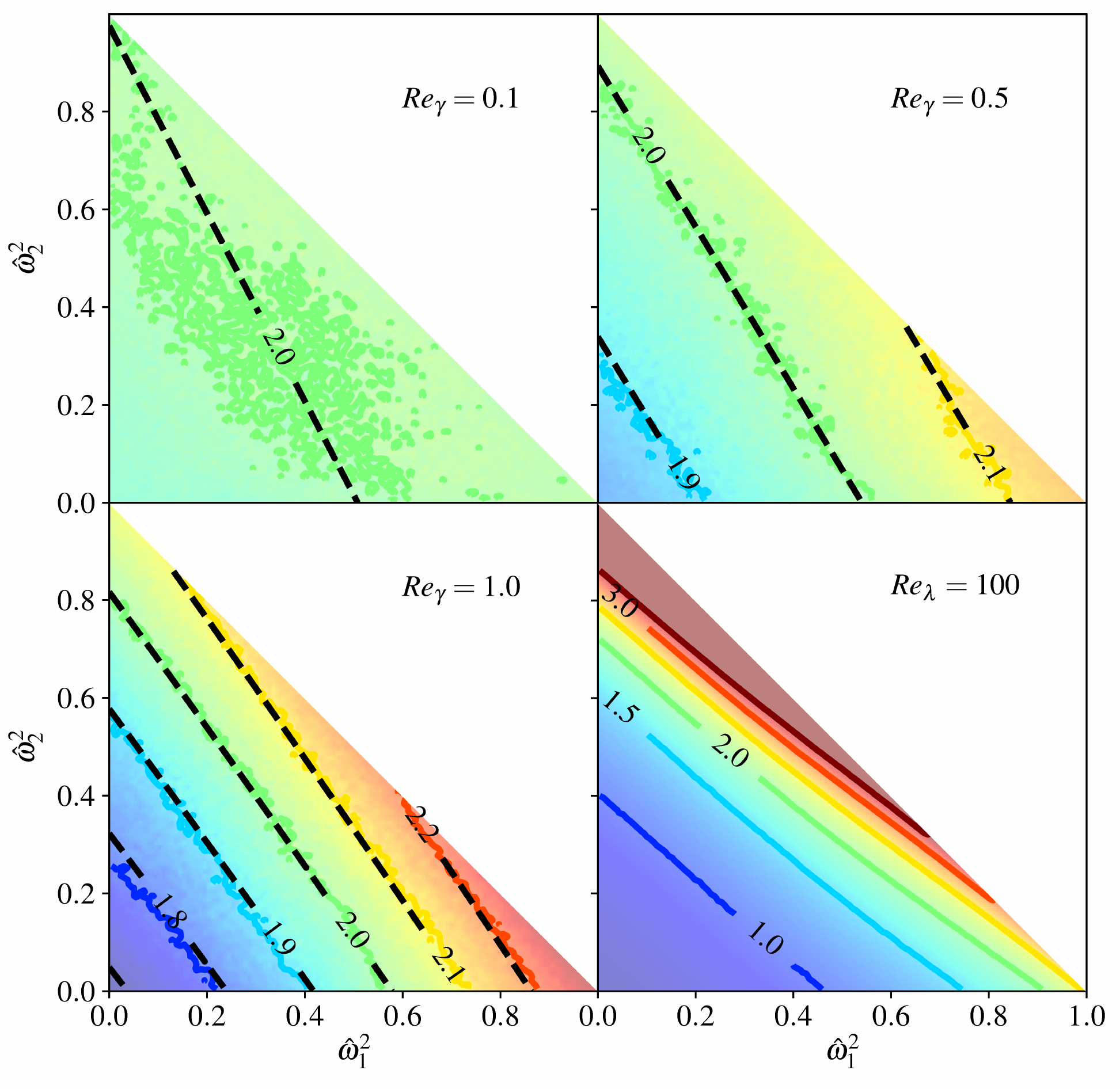}
  }
	\caption{PDF of the squared vorticity principal components weighted by the volume element, as in \eqref{eq_def_f_omg_weigh}, at various Reynolds numbers. $\hat{\omega}_1$ and $\hat{\omega}_2$ are the normalized vorticity components along the most extensional and intermediate strain-rate eigendirections, respectively. The colormap and coloured solid contours refer to the DNS results, while the black dotted lines refer to the corresponding analytic prediction \eqref{eq_pdf_omg_h2}. The numbers on the contours indicate the value of the PDF on that contour level.}
\label{fig_PDF_omg_h2}
\end{figure}

We now analyze the distribution of the vorticity components in the strain-rate eigenframe and explore the non-monotonic trend of the strain rate--vorticity preferential alignments with increasing Reynolds number observed in figure \ref{fig_mom_invs}(b).
As done for the strain rate in the previous section, we remove the kinematic effects due to the volume element $J_\omega$ \eqref{eq_vol_el_omg}, that arises from employing magnitude/orientation coordinates in the strain-rate eigenframe.
More specifically, we integrate out the strain rate and the vorticity magnitude from the velocity gradient PDF, according to equation \eqref{eq_ens_avg}, to obtain the marginal PDF of the normalized vorticity principal components
\begin{align}
\rho_{\hat{\omega}}(\hat{\omega}^2) = \int\de\omega J_\omega \int\de\tens{S}\, f(\inv\left(\tens{S},\vect{\omega})\right),
\label{eq_def_rho_omg}
\end{align}
where the integration with respect to the strain rate is defined in \eqref{eq_ens_avg_str}. We then weigh the PDF of the alignments $\rho_{\hat{\omega}}$ by the denominator of the volume element $J_\omega$ \eqref{eq_vol_el_omg}, which does not depend on the vorticity magnitude and can be pulled out of the integral in \eqref{eq_def_rho_omg}, thus yielding
\begin{align}
f_{\hat{\omega}}(\hat{\omega}^2) =
\mathcal{N}\sqrt{\hat{\omega}_1^2\hat{\omega}_2^2\hat{\omega}_3^2}
\rho_{\hat{\omega}}(\hat{\omega}^2)
\label{eq_def_f_omg_weigh}
\end{align}
where $\mathcal{N}$ is a normalization factor such that $\int\de\hat{\omega}_1^2\de\hat{\omega}_2^2f_{\hat   {\omega}}=1$.
The advantage of such re-weighting of the PDF is that it removes the complicated features of the volume element $J_\omega$, resulting in visually simple trends of the alignment PDF, as shown in figure \ref{fig_PDF_omg_h2}.

In figure \ref{fig_PDF_omg_h2} we compare  the re-weighted PDF of the alignments $f_{\hat{\omega}}$ from DNS (coloured lines) with our low-Reynolds number model prediction (black dotted lines). The asymptotic prediction of $f_{\hat{\omega}}$ follows from integrating out the strain rate and vorticity magnitude from the asymptotic solution \eqref{eq_f_exp_Re} and it reads
\begin{align}
&f_{\hat{\omega}}\left(\hat{\omega}^2\right) = 2 - \Rey_\gamma\frac{18 \sqrt{30} S_{3} }{5 \sqrt{\pi}}\left(2 \hat{\omega}_{1}^2 + \hat{\omega}_{2}^2 - 1\right)
 +\Rey_\gamma^2\Bigg(228 S_{3}^{2} \hat{\omega}_{1}^4 + 228 S_{3}^{2} \hat{\omega}_{1}^2 \hat{\omega}_{2}^2 +\nonumber\\
 &- 228 S_{3}^{2} \hat{\omega}_{1}^2 + 66 S_{3}^{2} \hat{\omega}_{2}^4 - \frac{59856 S_{3}^{2} \hat{\omega}_{2}^2}{245} + \frac{21912 S_{3}^{2}}{245} - \frac{10692 X_{5} \hat{\omega}_{2}^2}{49} + \frac{3564 X_{5}}{49}\Bigg),
\label{eq_pdf_omg_h2}
\end{align}
where, the vorticity components are ordered according to the corresponding strain-rate eigenvalue, with $\lambda_1>\lambda_2>\lambda_3$, and two of the normalized vorticity principal components suffice to parameterize the alignment PDF since $\sum_i\hat{\omega}_i^2=1$.

At very small $\Rey_\gamma$, the alignment distribution in figure \ref{fig_PDF_omg_h2} is close to random uniform, corresponding to an almost constant $f_{\hat{\omega}}$ and to the lack of preferential alignments. At low Reynolds numbers, the probability density re-weighted by the volume element, $f_{\hat{\omega}}$, displays almost straight contours, which have a slope such that the PDF takes larger values where $\hat{\omega}_1^2$ is large. The analytic solution \eqref{eq_pdf_omg_h2} shows that the contours at first order in the Reynolds number consist indeed of straight lines with slope $-2$ in the $\hat{\omega}_1^2$--$\hat{\omega}_2^2$ plane.
This results in the preferential alignment between the vorticity and the most extensional strain-rate eigendirection, as previously observed in figure \ref{fig_mom_invs}. Moreover, the first-order correction enhances the probability of regions in which $\hat{\omega}_1^2+\hat{\omega}_2^2$ is large since $S_3<0$. This shows a connection between the negative skewness of the strain-rate statistics and the lack of alignment between the vorticity and the most compressional strain-rate direction.
As the Reynolds number increases, the contours of the alignment PDF \ref{fig_PDF_omg_h2} tilt toward preferentially large $\hat{\omega}_2^2$, that is a slope larger than $-1$ in the $\hat{\omega}_1^2$--$\hat{\omega}_2^2$ plane. The contours remain approximately straight at low Reynolds numbers, with mild variations of the re-weighted PDF magnitude on its support. The analytic solution \eqref{eq_pdf_omg_h2} quantitatively captures  the PDF and the tilting of the contours with increasing $\Rey_\gamma$, the tilting stemming from second-order terms in $\Rey_\gamma$.
At large Reynolds numbers, the PDF varies more rapidly on its support, and its contours visually deviate from straight lines. However, the low- and high-Reynolds number re-weighted PDF of the alignments still look remarkably similar, up to a tilting of the contours.

\begin{figure}
  \centerline{
  \includegraphics[width=\textwidth]{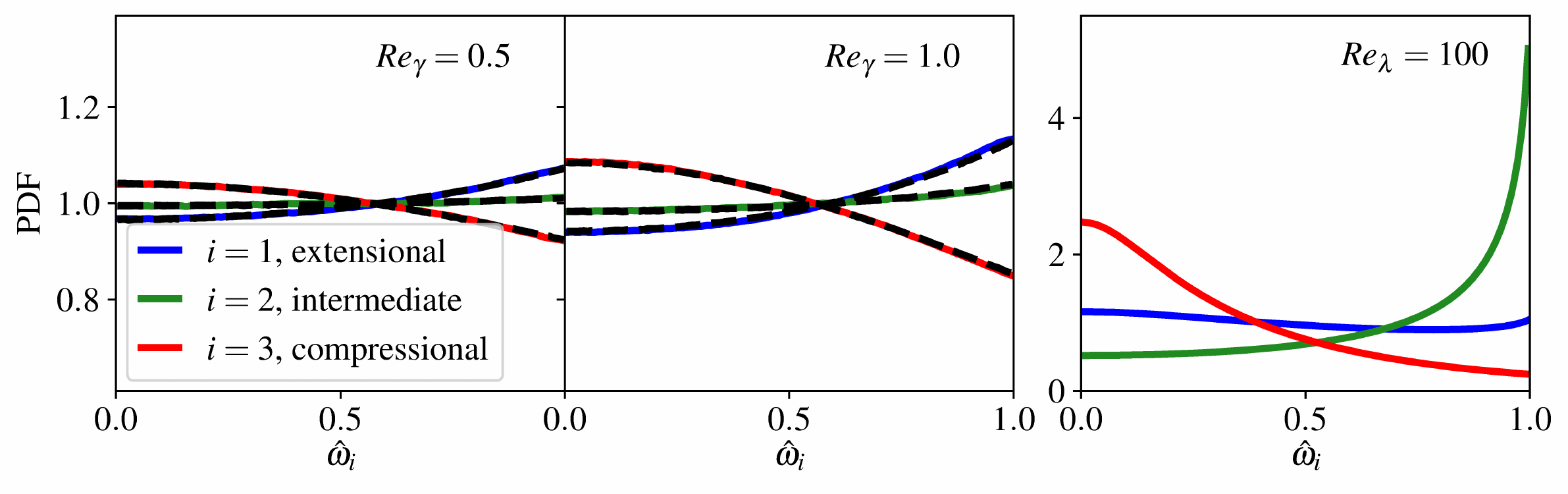}
  }
	\caption{Vorticity-strain rate alignments quantified by the PDF of the ordered vorticity principal components at various Reynolds numbers. The solid coloured lines refer to the DNS, while the black dotted lines are from the numerical solution of our low-Reynolds number model \eqref{eq_Langevin_num}.}
\label{fig_PDF_omg_align}
\end{figure}

The tilting of the contours of the normalized vorticity principal components PDF toward preferentially large $\hat{\omega}_2^2$ reflects in changes of the marginal PDFs of the cosines $\hat{\omega}_i=|\vect{v}_i\bcdot\vect{\omega}|$, where $\vect{v}_i$ are the ordered strain-rate eigenvectors.
In this case, we could not compute the analytic approximation to the marginal PDF of the alignments due to the technical difficulty introduced by the volume element  $1/|\hat{\omega}_1\hat{\omega}_2\hat{\omega}_3|$ featured in the integrals \eqref{eq_ens_avg}. Instead, we compute the PDF of the alignments by numerically solving the Langevin equation associated with the FPE \eqref{eq_FPE_Cart} for an ensemble of velocity gradients
\begin{align}
\frac{\de \tens{A}}{\de t} &= -\bigg[
\left(1 - \Rey_\gamma^2\delta_1\right)\tens{S}  +
\left(1 - \Rey_\gamma^2\delta_2\right)\tens{W} + \Rey_\gamma\left(\frac{5}{7}-\delta_3\right)\widetilde{\tens{SS}} +
\Rey_\gamma\gamma_4\left(\tens{SW}-\tens{WS}\right)+ \nonumber\\
&+\Rey_\gamma\left(1-\delta_5\right)\left(\tens{SW}+\tens{WS}\right) +
\Rey_\gamma\left(\frac{3}{5}-\delta_6\right)\widetilde{\tens{WW}}\bigg] + \sigma \bm{\Gamma}
\label{eq_Langevin_num}
\end{align}
where $\bm{\Gamma}$ is a white-in-time tensorial noise which has the same (single-point/single-time) statistics of the forcing $\bnabla \vect{F}$ \eqref{eq_f_correl_phys}.

Figure \ref{fig_PDF_omg_align} shows the PDF of the normalized vorticity principal components $\hat{\omega}_i$, obtained from DNS and from numerical integration of our low-Reynolds number model \eqref{eq_Langevin_num}.
At very small $\Rey_\gamma$, there are no preferential alignments, and the distribution of the normalized principal components is random uniform. As $\Rey_\gamma$ increases, the vorticity preferentially aligns first with the most extensional strain-rate principal direction while avoiding alignment with the compressional direction.
Since the spinning of the fluid element causes compression along the vorticity direction \citep{Carbone2020b}, the lack of alignment between vorticity and the contracting direction hinders the growth of the most compressional velocity gradients.
The transient alignment between the vorticity and the extensional strain-rate eigenvector lasts till $\Rey_\gamma \approx 5$, and then the well-known preferential alignment with the intermediate eigenvector settles \citep{Ashurst1987}.

\subsection{Velocity gradient principal invariants}

Finally, we focus on the joint PDF of the velocity gradient principal invariants, namely
\begin{align}
\mathscr{Q} = -\frac{1}{2}\left(\inv_1+\inv_2\right), 
&&
\mathscr{R} = -\frac{1}{3}\left(\inv_3+3\inv_4\right),
\label{eq_def_QR}
\end{align}
that is a hallmark in the study of the turbulent velocity gradient dynamics \citep{Meneveau2011}.
The volume element characterizing the $\mathscr{R}$--$\mathscr{Q}$ space prevented us from analytic integration of the full PDF \eqref{eq_f_exp_Re} to obtain the marginal PDF of the principal invariants. As for the strain rate--vorticity alignments presented above, we obtain the marginal $\mathscr{R}$--$\mathscr{Q}$ PDF by numerically solving the Langevin equation \eqref{eq_Langevin_num}.

\begin{figure}
  \centerline{
  \includegraphics[width=.9\textwidth]{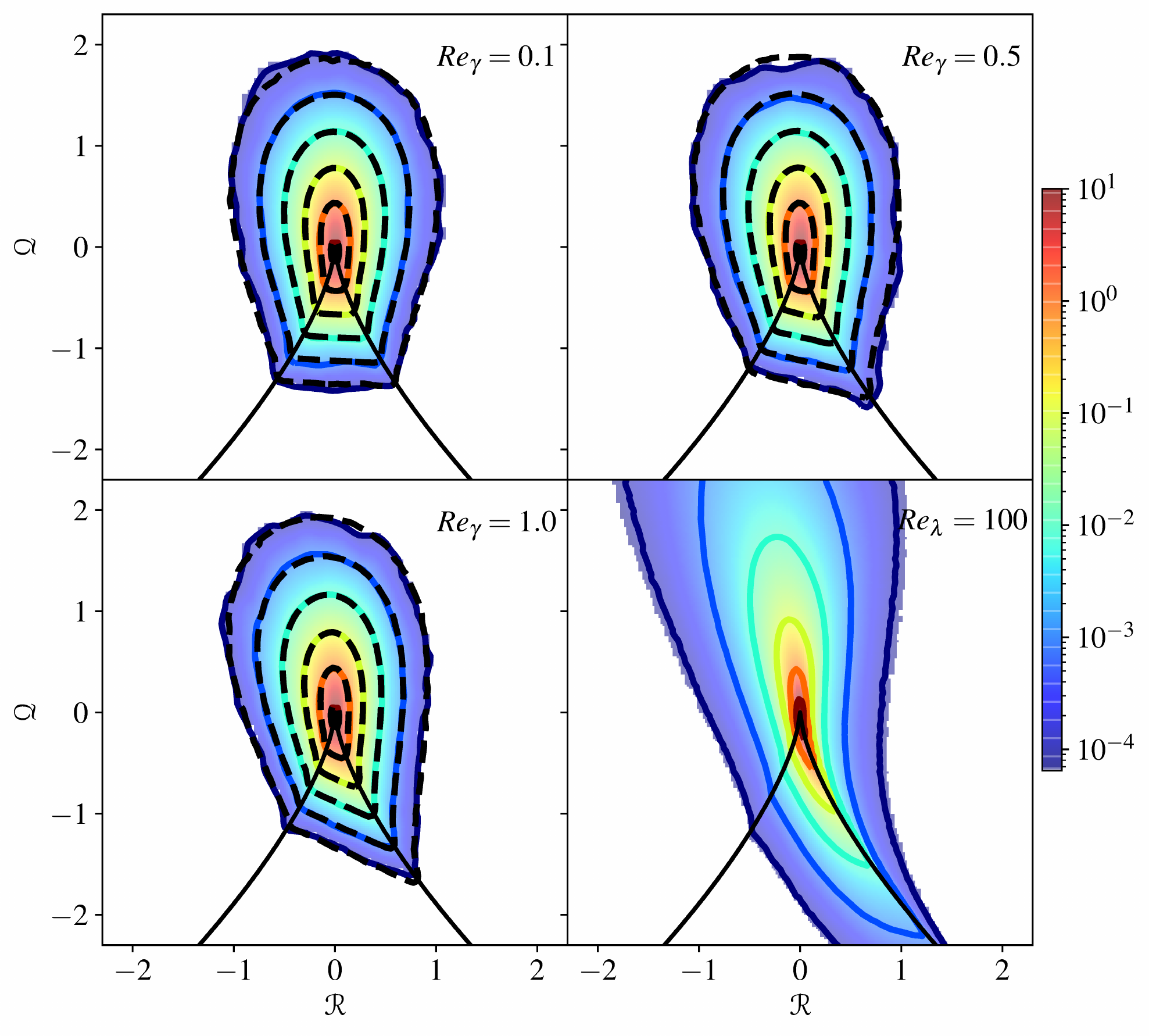}
  }
  \caption{Probability density of the velocity gradient principal invariants \eqref{eq_def_QR} in $\log_{10}$ scale. The colormap and coloured solid contours refer to the DNS, while the black dotted lines are from the numerical solution of our low-Reynolds number model \eqref{eq_Langevin_num}.}
\label{fig_PDF_RQ}
\end{figure}

The $\mathscr{R}$--$\mathscr{Q}$ PDF entangles all the information on the strain rate and vorticity in a somewhat complicated way. Indeed, integrating out the eigenvectors of $\tens{A}$, in order to get the marginal $\mathscr{R}$--$\mathscr{Q}$ PDF, does not immediately relate to integrating out rotations of the reference frame, as it is instead for the strain rate. Also, the $\mathscr{R}$--$\mathscr{Q}$ PDF features two kinematically different regions, one below the Vieillefosse line \citep{Vieillefosse1982} corresponding to real velocity gradient eigenvalues and one above it corresponding to complex eigenvalues. This topological difference causes edges in the PDF of the principal invariants, even though the gradient statistics are smooth.

Nonetheless, the $\mathscr{R}$--$\mathscr{Q}$ PDF displays a distinguishing mark of the velocity gradient statistics: the classical teardrop shape, elongated towards the right Vieillefosse tail \citep{Vieillefosse1982}, as shown in figure \ref{fig_PDF_RQ}. At infinitesimally small Reynolds number, the PDF takes the well-known Gaussian shape, symmetric along the $\mathscr{R}$ axis due to invariance of equations \eqref{eq_Langevin_A} under the sign flipping $\tens{A}\rightarrow -\tens{A}$ \citep{Meneveau2011}.
As $\Rey_\gamma$ increases, the characteristic teardrop shape arises. Our model quantitatively captures the onset of the skewness along the right Vieillefosse tail, which persists at higher Reynolds numbers. However, the similarity between the low-Reynolds and high-Reynolds number PDFs is only qualitative, the large Reynolds number PDF displaying significantly more elongated tails. This is also because the principal invariants are moments of the velocity gradient of degrees two and three, and tiny tails in the strain-rate eigenvalues and vorticity components distributions result in pronounced tails of the $\mathscr{R}$--$\mathscr{Q}$ PDF.

\section{Comparison of  multi-time statistics and velocity gradient dynamics}
\label{sec_trajectories}

In the previous section, we showed how the asymptotic solution \eqref{eq_f_exp_Re} quantitatively captures the single-time/single-point velocity gradient statistics obtained from direct numerical simulations up to $\Rey_\gamma\simeq 1$, with a qualitative agreement till $\Rey_\gamma\simeq 5$.
Now, we investigate whether our model can also reproduce the full Lagrangian dynamics of the velocity gradients, as characterized by velocity gradient sample realizations and time correlations.

\subsection{Model coefficients from the DNS}

To compare the velocity gradient time series generated by the DNS with our low-Reynolds number model predictions, we first need to fully specify all the model coefficients \eqref{eq_model_coeff}.
Indeed, in section \ref{sec_theory}, we have identified two gauge coefficients, $\gamma_4$ and $\zeta_5$, which only affect multi-time statistics while leaving the single-time PDFs unchanged. 
We compute the model coefficients \eqref{eq_model_coeff} directly from the DNS data and compare them with their analytic predictions. This allows us fitting the two gauge coefficients, $\gamma_4$ and $\zeta_5$, from DNS data, thus enabling our low-Reynolds number model to capture the velocity gradient temporal dynamics.

From the direct numerical simulation of low-Reynolds random flows, we have access to space-time realizations of the pressure Hessian and viscous Laplacian, i.e., the unclosed terms in our single-time/single-point modelling approach. Therefore, we can compute from DNS the averages of the unclosed terms conditional on the local velocity gradient via tensor function representation theory \citep{Leppin2020,Carbone2021}
\begin{subequations}
\begin{align}
\cavg{\widetilde{\tens{H}}(\vect{x},t)}{\tens{A}} &= \sum_{n=1}^8 h_n \tens{B}^n,
\label{eq_tens_repr_H}
\\
\cavg{\nabla^2\tens{A}(\vect{x},t)}{\tens{A}} + \tens{A} &= \Rey_\gamma^2\sum_{n=1}^2 \delta_n \tens{B}^n + \Rey_\gamma\sum_{n=3}^8 \delta_n \tens{B}^n.
\label{eq_tens_repr_V}
\end{align}
\label{eq_tens_repr_H_V}
\end{subequations}
Here $\tens{B}^n$ are the basis tensors \eqref{eq_def_BT}, $h_n$ in \eqref{eq_tens_repr_H} are the conditional pressure Hessian components, and $\delta_n$ in \eqref{eq_tens_repr_V} are the components of the viscous corrections.
To be consistent with our low-Reynolds number model formulation, here we use the same representation of the pressure Hessian and viscous Laplacian that we employed to derive the model coefficients \eqref{eq_model_coeff}. From the viscous Laplacian \eqref{eq_tens_repr_V}, we subtract the linear damping part, $-\tens{A}$, and then we introduce second-order corrections in the Reynolds number proportional to the first two coefficients $\delta_1$ and $\delta_2$, and first-order corrections proportional to the other basis tensors.

To extract the coefficients $h_n$ and $\delta_i$ from the DNS data, we use the following property of any conditional average (we illustrate this only for the anisotropic pressure Hessian)
\begin{align}
\avg{\cavg{\widetilde{H}_{ij}(\vect{x},t)}{\tens{A}} T_{ij}(\tens{A})} = \avg{\widetilde{H}_{ij}(\vect{x},t) T_{ij}\left(\tens{A}(\vect{x},t)\right)}
\label{eq_cond_avg_prop}
\end{align}
where $\tens{T}(\tens{A})$ is a tensor function of the gradient only.
At most, the model coefficients depend on all the five invariants \eqref{eq_def_inv}, while our main modelling hypothesis is that we limit ourselves to constant coefficients.
This hypothesis is exact for the conditional pressure Hessian and the zeroth-order viscous Laplacian at very low Reynolds number, while it constitutes an approximation for higher-order corrections.

\begin{figure}
  \centerline{
  \begin{overpic}[width=\textwidth]{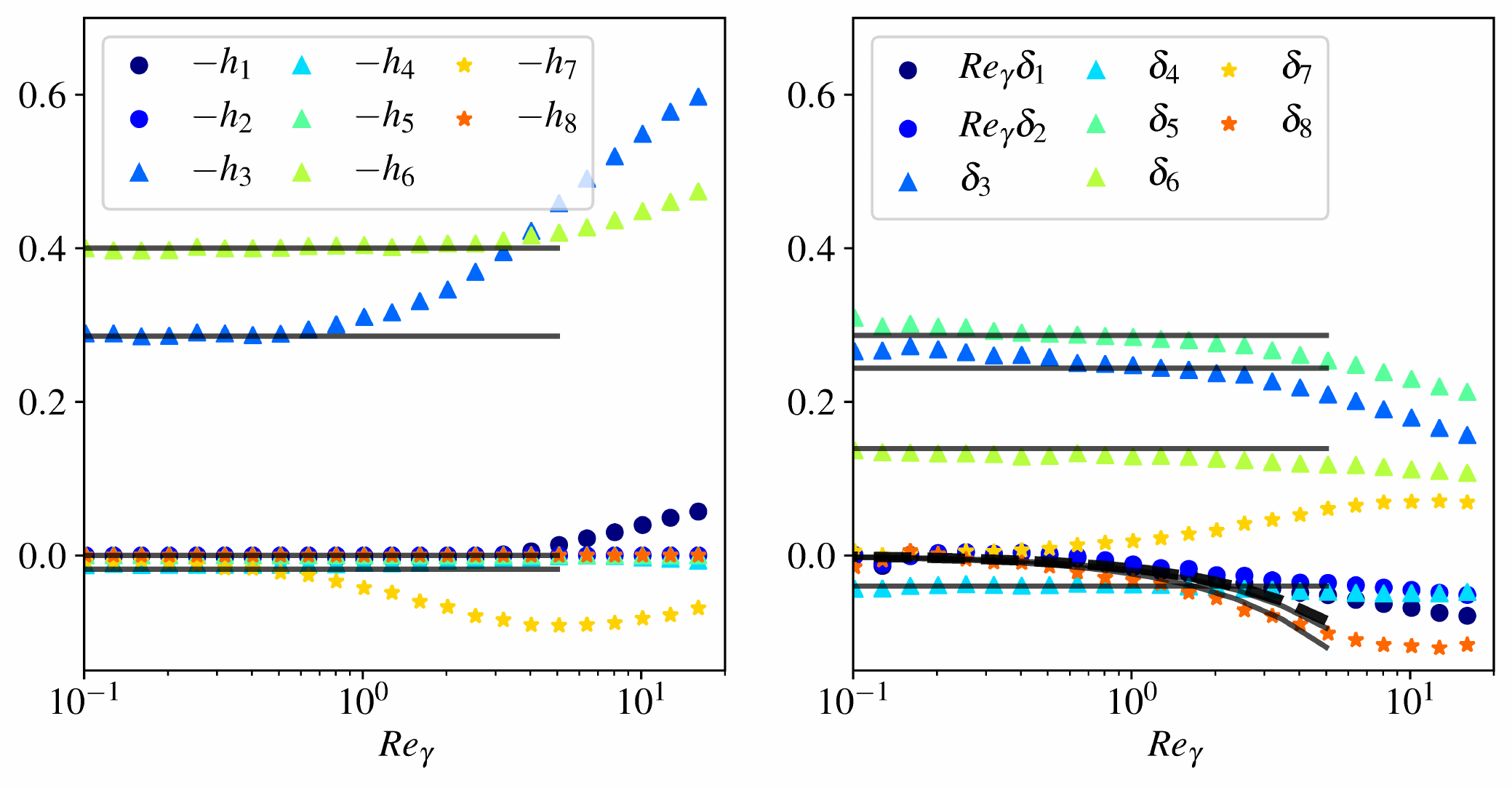}
\put(26,52){(\textit{a})}
\put(75,52){(\textit{b})}
  \end{overpic}
  }
	\caption{Components of the conditional anisotropic pressure Hessian $h_n$ (a) and (b) components of the conditional viscous Laplacian $\delta_n$ (see equation  \eqref{eq_tens_repr_H_V} for their definition), as functions of the Reynolds number $\Rey_\gamma$. The coloured points indicate the coefficients computed from DNS using \eqref{eq_lin_sys_delta}, with the point types differentiating the order of the corresponding basis tensor \eqref{eq_def_BT}. The gray transparent lines are from the analytic prediction \eqref{eq_model_coeff}, after fitting the gauge parameters $\gamma_4$ and $\zeta_5$. The model coefficients $\delta_7$ and $\delta_8$ are set to zero and not shown here.}
\label{fig_model_coeff}
\end{figure}

Thanks to the property \eqref{eq_cond_avg_prop}, we obtain a linear system for the constant model coefficients by taking the double contraction between the tensor representations \eqref{eq_tens_repr_H_V} and the basis tensors \eqref{eq_def_BT} and then averaging
\begin{subequations}
\begin{align}
\sum_{n=1}^8 \avg{Z^{mn}} h_n &= \avg{\widetilde{H}_{ij}(\vect{x},t) B^m_{ij}\left(\tens{A}(\vect{x},t)\right)},
\\
\Rey_\gamma^2 \sum_{n=1}^2 \avg{Z^{mn}}\delta_n + \Rey_\gamma \sum_{n=3}^8 \avg{Z^{mn}}\delta_n &= \avg{\left(\nabla^2 A_{ij}(\vect{x},t) + \tens{A}(\vect{x},t)\right) B^m_{ij}\left(\tens{A}(\vect{x},t)\right)},
\end{align}
\label{eq_lin_sys_delta}
\end{subequations}
where $Z^{mn}$ is the metric tensor \eqref{eq_def_met_tens}. Solving the linear system \eqref{eq_lin_sys_delta} yields the coefficients $h_n$ and $\delta_n$, plotted in figure \ref{fig_model_coeff} as functions of the Reynolds number $\Rey_\gamma$.

Figure \ref{fig_model_coeff}(a) shows that the pressure Hessian coefficients $h_3$ and $h_6$ (relative to the basis tensors $\widetilde{\tens{SS}}$ and $\widetilde{\tens{WW}}$ respectively) constitute the dominant contribution at low Reynolds numbers, with $h_3\to -2/7$ and $h_6\to -2/5$ at vanishingly small $\Rey_\gamma$ \citep{Wilczek2014}. The coefficient $h_7$ is largely compensated by the viscous part, which is beneficial for modelling since we hypothesized that the third-order basis tensors do not contribute to the gradient dynamics.
At $\Rey_\gamma\simeq 5$,
the coefficient $h_3$, which mitigates the strain-rate self-amplification, becomes larger in magnitude than the coefficient $h_7$, which instead hinders the centrifugal forces due to the rotation of the fluid element. At the same Reynolds threshold, the conditional pressure Hessian component along $\tens{B}^1\equiv \tens{S}$ starts increasing.

Figure \ref{fig_model_coeff}(b) shows that the corrections to the viscous linear damping encoded in $\delta_1$ and $\delta_2$ are moderate, and the model qualitatively captures the growth in magnitude of $\delta_1$ and $\delta_2$, which slowly become more negative as the Reynolds number increases. The fact that the coefficients $\Rey_\gamma\delta_1$ and $\Rey_\gamma\delta_2$ remain small at small Reynolds numbers justifies the assumption of quadratic (or, at least, higher-order than linear) corrections to the viscous damping.
As for the pressure Hessian, also for the viscous stress the dominant contributions come from the second-order basis tensors, $\tens{B}^3$ to $\tens{B}^6$.
The cubic terms are subleading at small Reynolds number and become relevant only at $\Rey_\gamma\gtrapprox 1$. While the contributions to the symmetric part $\tens{B}^7$ from the pressure Hessian and viscous stress compensate, the viscous anti-symmetric part proportional to $\tens{B}^8$ becomes relevant when $\Rey_\gamma\gtrapprox 1$. This indicates that velocity gradient models at higher Reynolds numbers would need to take into account those third-order basis tensors.

Remarkably, the coefficient $\delta_3$, relative to $\tens{B}^3\equiv \widetilde{\tens{SS}}$, is uniquely determined by the skewness growth rate at small Reynolds $S_3$, as shown by the asymptotic prediction \eqref{eq_model_coeff}. The strong relationship between the onset of skewness in the gradient statistics and the coefficient of $\tens{B}^3$ has been noticed before in the numerical experiments of \cite{Leppin2020}, and here it is shown analytically.
Furthermore, both $S_3$ and $\delta_3$ simultaneously match their DNS values (cf.~figure \ref{fig_mom_invs}(a) and \ref{fig_model_coeff}(b)), which have been obtained in two independent ways ($S_3$ by looking at $\avg{\inv_3}$ as a function of $\Rey_\gamma$, and $\delta_3$ by computing the conditional viscous Laplacian). This supports the consistency of our modelling approach.

Finally, the analytic expressions of the coefficients $\delta_4$, $\delta_5$ and $\delta_6$ in \eqref{eq_model_coeff}, feature the gauge terms $\gamma_4$ and $\zeta_5$. While the single-time statistics are independent of those two terms, the time correlations are affected. We fit the values of those gauge terms from DNS data, and with
\begin{align}
\gamma_4\simeq -0.022, && \zeta_5\simeq 0.013,
\end{align}
the analytic prediction \eqref{eq_model_coeff} matches well the coefficients $\delta_n$ independently computed from DNS, as in figure \ref{fig_model_coeff}(b).

\subsection{Lagrangian trajectories}

Now that we have fully determined the model coefficients, we analyze the velocity gradient along fluid particle trajectories generated by our model and by DNS.
The model coefficients necessary to reproduce the single-time velocity gradient statistics have been determined analytically. However, fitting DNS data is necessary to set the gauge terms in \eqref{eq_model_coeff}, affecting the velocity gradient time correlations.
The Langevin equation \eqref{eq_Langevin_num} and the associated Fokker-Planck equation \eqref{eq_FPE_Cart} are designed to match single-time/single-point statistics, and the results can be interpreted in either an Eulerian or Lagrangian sense. In the following, we assume a Lagrangian viewpoint.

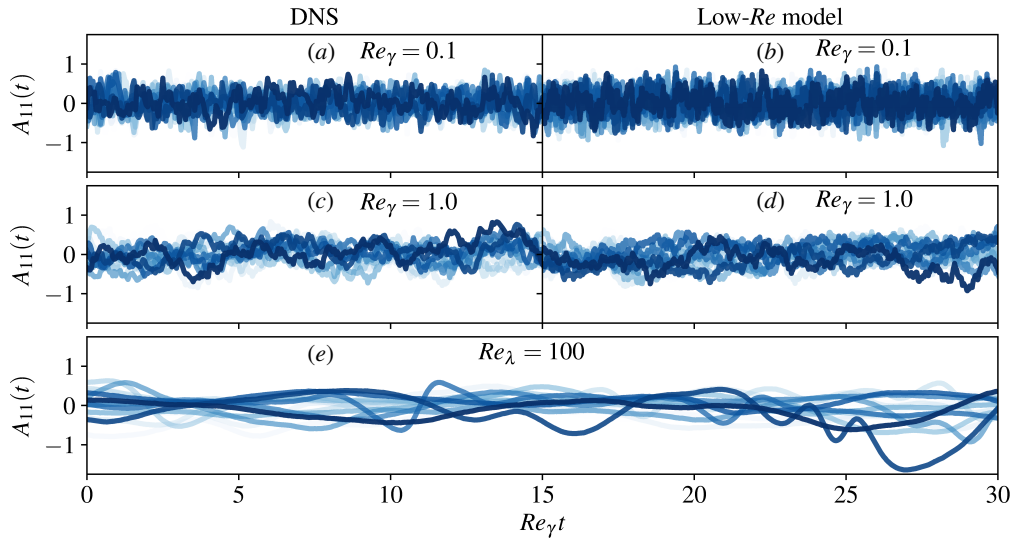
\begin{figure}
  \centerline{
\begin{overpic}[width=\textwidth]{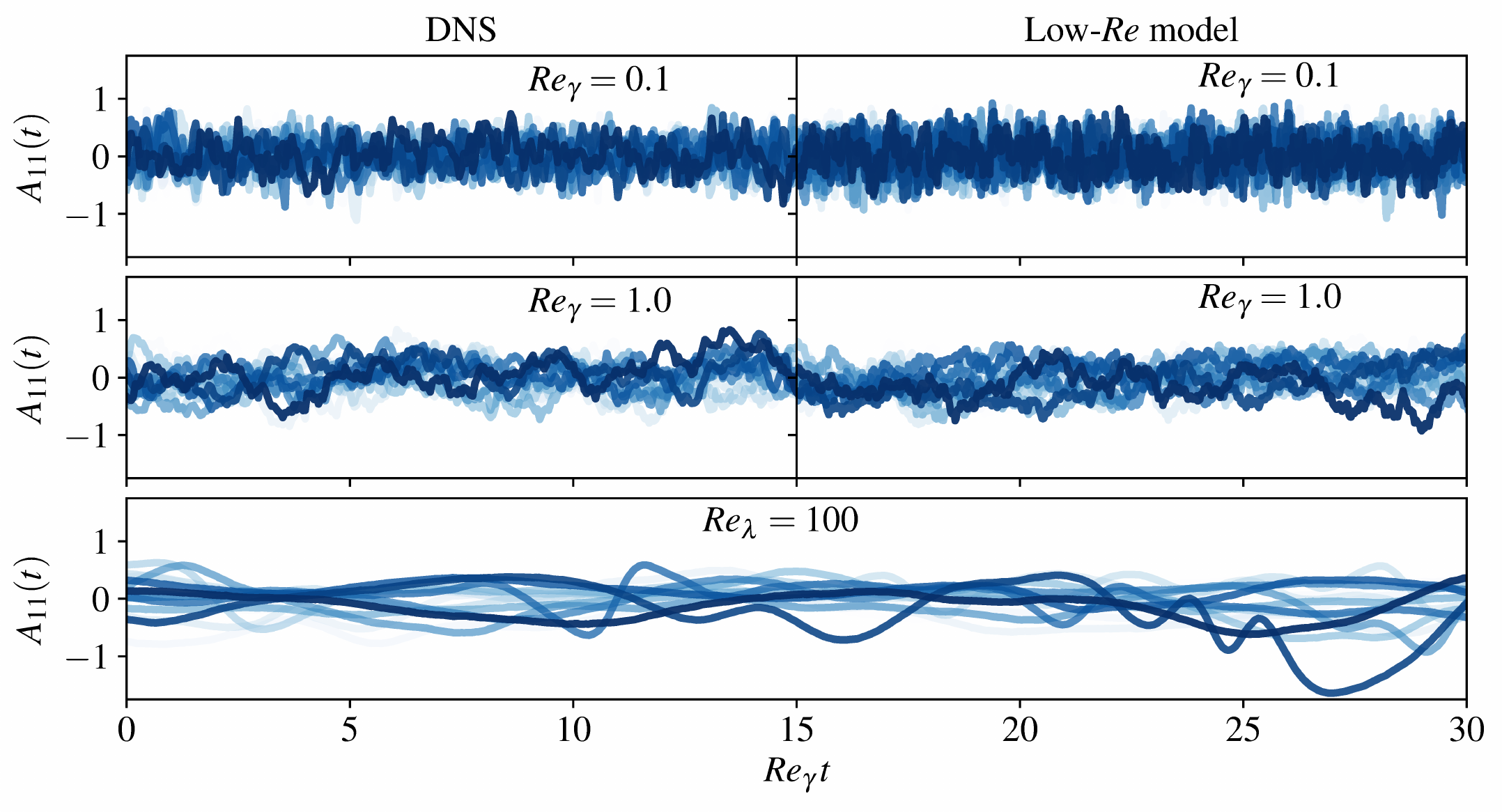}
\put(30,48){(\textit{a})}
\put(30,33.5){(\textit{c})}
\put(30,18.5){(\textit{e})}
\put(74,48){(\textit{b})}
\put(74,33.5){(\textit{d})}
\end{overpic}
}
	\caption{Time evolution of the velocity gradient longitudinal component at low Reynolds number, from DNS (a,c) and from our low-Reynolds number model \eqref{eq_Langevin_A} (b,d). The bottom panel (e) shows trajectories at moderately large Reynolds number. Curves at different transparency refer to various samples in the same simulation. The plots share the same horizontal axis, with the dimensional time normalized by the Kolmogorov time scale, $\bar{t}/\bar{\tau}_\eta=\Rey_\gamma\, t$.}
\label{fig_S11_trajectories}
\end{figure}

Figure \ref{fig_S11_trajectories} shows the longitudinal component of the strain rate along fluid-particle trajectories at various Reynolds numbers, from the model and from the DNS.
For a vanishingly small Reynolds number, the noise dominates and the realizations follow a tensorial Ornstein-Uhlenbeck process. As the Reynolds number increases, the effect of the random forcing weakens, time correlations establish and larger negative gradients persist. The symmetry about $A_{11}=0$ breaks as Reynolds increases, with large negative gradients being more likely, as evident especially at a large Reynolds number. The larger-Reynolds number trajectories observed on the time scale of a few Kolmogorov times appear smooth, indicating that the details of the stochastic forcing are concealed by the rich small-scale turbulent structure of the flow.
The velocity gradient realizations from our low-Reynolds number model are visually similar to the realizations from DNS up to $\Rey_\gamma=\ordof(1)$, and we are now going to quantify this similarity by comparing their time correlations.

\subsection{Time correlations from DNS and from the low-Reynolds number model}

\begin{figure}
  \centerline{
  \begin{overpic}[width=\textwidth]{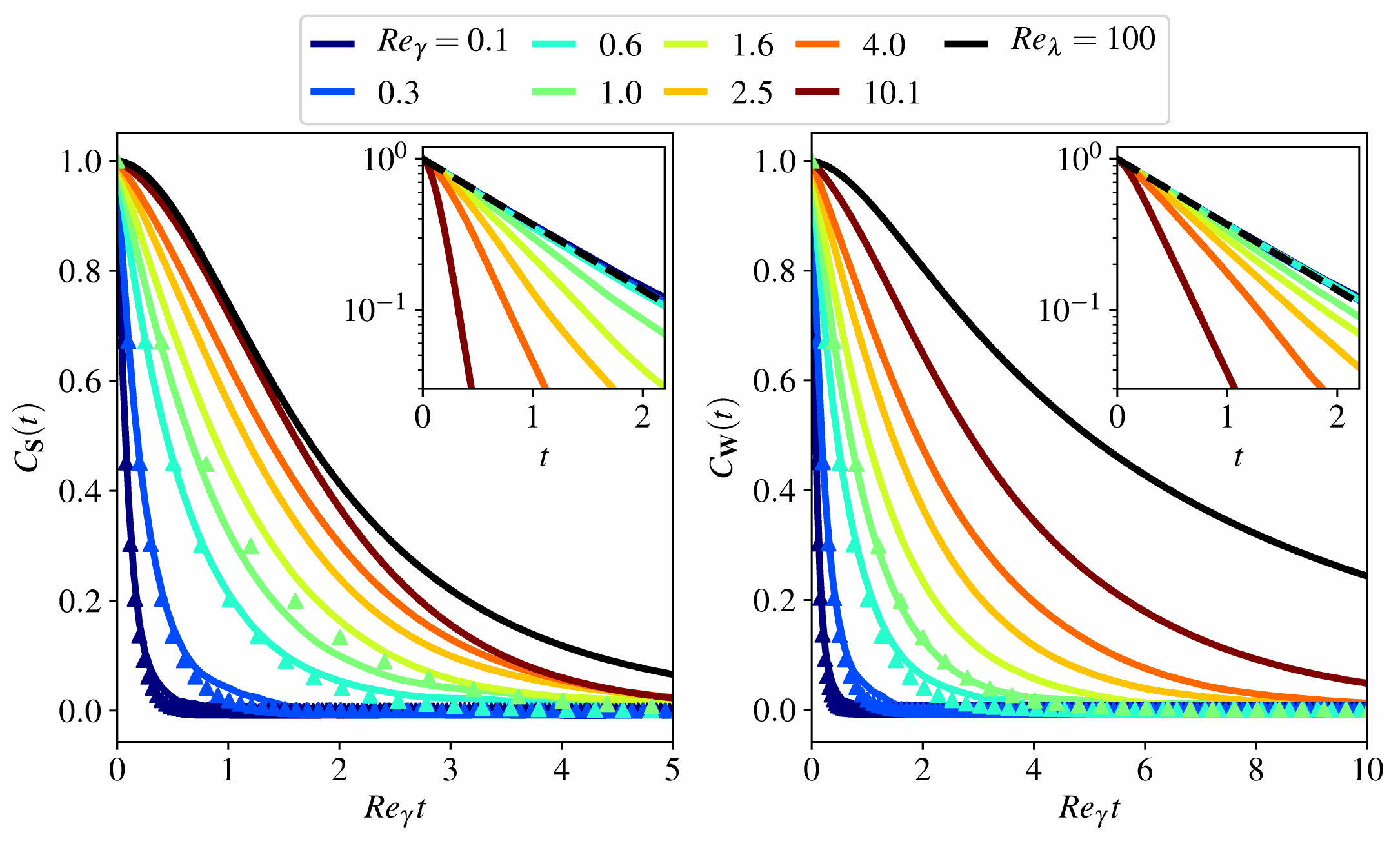}
\put(16,52){(\textit{a})}
\put(86,52){(\textit{b})}
\end{overpic}
}
	\caption{Normalized time correlations of the strain rate (a) and rotation rate (b), for various Reynolds numbers, as functions of the time lag normalized by the Kolmogorov time scale, $\bar{t}/\bar{\tau}_\eta=\Rey_\gamma\, t$. Solid lines are from the DNS, while the symbols refer to our low-Reynolds number model \eqref{eq_Langevin_A}. The insets show the correlations in a semi-logarithmic scale, as a function of the non-dimensional time lag, $t=(\bar{\nu}/\bar{\gamma}_0^2)\bar{t}$. The dashed black dotted lines in the insets indicate the expected zero-Reynolds number correlation, $C_{\bm{A}}= \exp(-t)$.}
\label{fig_correlations}
\end{figure}

In figure \ref{fig_correlations}, we compare the normalized time correlations of the strain and rotation rates
\begin{align}
	C_{\tens{S}}(t) = \frac{\avg{S_{ij}(t)S_{ij}(0)}}{\avg{S_{ij}(0)S_{ij}(0)}},
&&
	C_{\tens{W}}(t) = \frac{\avg{W_{ij}(t)W_{ij}(0)}}{\avg{W_{ij}(0)W_{ij}(0)}},
\end{align}
as obtained from the DNS and from our low-Reynolds number model \eqref{eq_Langevin_num}.
At very small $\Rey_\gamma$, the velocity gradient follows a linear Langevin equation so that the correlations decay exponentially in time, with characteristic time scale $\bar{\gamma}_0^2/\nu$. Therefore, in the non-dimensional time $t$ \eqref{def_nd_variables}, the correlations collapse at small Reynolds numbers, as evident from the insets in figure \ref{fig_correlations}.
However, as the Reynolds number increases, the correlations decay with a time scale shorter than $\bar{\gamma}_0^2/\nu$, indicating that the convective nonlinearities reduce time correlations.

Furthermore, the strain rate is correlated over shorter time scales, as in figure \ref{fig_correlations}(a), and the bulk of the correlations already establishes at $\Rey_\gamma = \ordof(10)$, that is just after the transition to turbulent configuration begins. Instead, the vorticity correlates over longer time scales and those long-lasting correlations only establish at relatively large Reynolds numbers.
The extreme vorticity intermittency and long-lasting correlations then constitute the main missing ingredients in low-Reynolds number random flows, as compared to turbulent flows. This is in agreement with recent observations \citep[e.g.][]{Ghira2022} that the most intense vortical structures fully develop only at very large Reynolds numbers.

Our model can quantitatively predict the velocity-gradient time correlations up to $\Rey_\gamma=\ordof(1)$, and to this end, the gauge terms discussed in the previous sections are crucial. Moreover, the model performs better on the vorticity time correlations than on the strain-rate correlations. The difficulty in predicting the strain-rate time correlations has been recently observed also in reduced-order models for the velocity gradient at high Reynolds number \citep{Leppin2020}.

\section{Conclusions}
\label{sec_conclusions}

We have derived a model for the velocity gradient in  low-Reynolds number flows governed by the stochastically forced Navier-Stokes equations. The model follows directly from a weak-coupling expansion of the Navier-Stokes equations, combined with homogeneity constraints on the velocity gradient moments, and it is asymptotically exact at small Reynolds numbers. It features the exact first-order corrections to the velocity gradient dynamics due to the pressure Hessian, while the viscous contributions are subject to modelling hypotheses. 

We have explored the rich statistical geometry of the velocity gradient tensor, by comparing the results from direct numerical simulation and our model.
The model can quantitatively predict the onset of non-Gaussianity in random low-Reynolds number flows up to unitary Reynolds number, and qualitative agreement persists till a transition to the velocity gradient configuration reminiscent of turbulence. Beyond this transition, the model predictions break down, as expected for a weak-coupling perturbative approach.
Single-time statistics follow from the model without inputs from the DNS.
On the other hand, capturing the time correlations requires fitting two gauge parameters from the DNS data.

The asymptotic solution to the model Fokker-Planck equation associated with our stochastic model is integrable analytically, thus characterizing the onset of skewness, alignments and intermittency at low Reynolds number through closed asymptotic expressions.
This allowed us to quantify the onset of hallmark turbulent features, like the alignments between the vorticity and the strain rate, the preferential configuration of the strain-rate eigenvalues and the characteristic teardrop-shape of the velocity gradient principal invariants PDF.

The methodology developed here may also be applicable to model the PDF of the velocity gradient starting from the Navier-Stokes equations at larger Reynolds numbers, for example by employing renormalization techniques \citep{Yakhot1986,Apolinario2019}.

\backsection[Acknowledgements]{We thank Gabriel B.~Apolinário, Andy D.~Bragg, Michele Iovieno and Alain Pumir for insightful comments on the manuscript. This work was supported by the Max Planck Society.
Computations were performed on the HPC system Cobra at the Max Planck Computing and Data Facility.
This project has received funding from the European Research Council (ERC) under the European Union's Horizon 2020 research and innovation programme (Grant agreement No. 101001081).}

\backsection[Declaration of interests]{The authors report no conflict of interest.}

\FloatBarrier
\appendix
\section{DNS details}
\label{app_DNS_details}

In this Appendix we summarize the setup of the direct numerical simulations. We employ dimensional variables, denoted by a bar, the unit of measure being (arbitrary) code units. This is because the DNS plays the role of real-world experiments in our setup, and from those experiments we determine the reference length $\bar{\gamma}_0$, such that the relation for the conditional Laplacian \eqref{eq_Lapl_Gauss} holds. Once we determine $\bar{\gamma}_0$ and the dimensional dissipation rate $\bar{\varepsilon}$ of turbulent kinetic energy, we can re-scale all the dimensional variables \eqref{def_nd_variables} to write up the low-Reynolds number model in its non-dimensional form (\ref{eq_FPE_Cart}, \ref{eq_Langevin_num}).

We perform direct numerical simulations of incompressible flows governed by the Navier-Stokes equations stirred through a Gaussian random forcing by means of a standard Fourier pseudo-spectral method, described in \citep[e.g.][]{Carbone2019}. The code solves the incompressible Navier-Stokes equations on a tri-periodic cubic domain in the form
\begin{align}
\bar{k}_i\widehat{u}_i = 0, &&
(\partial_{\bar{t}} + \bar{\nu}\bar{k}^2) \widehat{u}_i + \imag P_{ij} \bar{k}_l \mathcal{F}\left[\bar{u}_j \bar{u}_l\right] 
=
\sqrt{\bar{\nu}} \bar{\sigma}_0 \bar{k} \widehat{F}_i,
\label{eq_NS_spectral}
\end{align}
where $\mathcal{F}$ indicates the spatial Fourier transform, $P_{ij}(\vect{k})=\delta_{ij}-k_ik_j/k^2$ is the projection tensor on the plane orthogonal to the wavevector $\bar{\vect{k}}$, $\bar{k}=\|\bar{\vect{k}}\|$ is the wavevector norm, $\imag$ is the imaginary unit, and
$\widehat{\vect{u}}$ is the transformed velocity field,
\begin{align}
\widehat{\vect{u}} (\bar{\vect{k}},\bar{t}) = 
\frac{1}{\bar{L}^3}\int \de \bar{\vect{x}} \bar{\vect{u}} (\bar{\vect{x}},\bar{t})\exp(-\imag \bar{\vect{k}}\bcdot\bar{\vect{x}}).
\end{align}
The domain length is $\bar{L} = 2\pi$ in code units, and the integer wavevectors components range in the interval $-N/2<\bar{k}_i<N/2$, where $N$ is the number of resolved Fourier modes in each direction \citep{Canuto2006}.
The grid spacing in Fourier space in each direction is $\Delta \bar{k}=2\pi/\bar{L}$ (and $\Delta \bar{k}=1$ in code units).
The low-Reynolds number simulations resolve $64^3$ Fourier modes, with spatial resolutions ranging from $\bar{k}_{\max}\bar{\eta}\simeq 40$ to $\bar{k}_{\max}\bar{\eta}\simeq 5$ with increasing Reynolds number $\Rey_\gamma$. Here $\bar{\eta}=(\bar{\nu}^3/\bar{\varepsilon})^{1/4}$ is the Kolmogorov length scale.
The  numerical simulation at larger Reynolds number resolves $512^3$ Fourier modes with $\bar{k}_{\max}\bar{\eta}\simeq 3$ and Taylor Reynolds number $\Rey_\lambda\approx 100$.
The aliasing error introduced by the non-linear convective term is removed through a $3/2$ rule \citep{Canuto2006}, and the time stepping consists of a second-order stochastic Runge-Kutta algorithm \citep{Honeycutt1992}.

In all the simulations, the complex Gaussian random forcing in \eqref{eq_NS_spectral} is limited to low wavenumbers, and it has the form
\begin{align}
\hat{F}_i(\bar{\vect{k}},t) = 
(\Delta \bar{k})^2 \sum_l
P_{ij}(\vect{k}_l) \left[
\left(\bar{a}_j^l + \imag \bar{b}_j^l\right) \delta\left(\bar{\vect{k}} - \bar{\vect{k}}_l\right)
+
\left(\bar{a}_j^l - \imag \bar{b}_j^l\right) \delta\left(\bar{\vect{k}} + \bar{\vect{k}}_l\right)
\right]
\label{eq_def_forcing}
\end{align}
where $\bar{\vect{a}}^l$ and $\bar{\vect{b}}^l$ are real, vectorial, white-in-time Gaussian random processes with zero mean and unitary variance, $\langle\bar{a}_i^l(\bar{t})\bar{a}_j^m(\bar{t}')\rangle=
\delta_{lm}\delta_{ij}\delta(\bar{t}-\bar{t}')$. The sum in \eqref{eq_def_forcing} is extended to the forced wavenumbers $\bar{\vect{k}}_l$, $1\le\|\bar{\vect{k}}_l\|< \bar{K}$, with $\bar{K}=\sqrt{7}$ (in code units).
The noise amplitude in \eqref{eq_NS_spectral} is proportional to the wavenumber, so that the kinetic energy spectrum $E_k$ is proportional to $\|\vect{k}\|^2$ at low wavenumbers.
For $\Rey_\gamma\to 0$, equation \eqref{eq_NS_spectral} tends to an Ornstein-Uhlenbeck process in which the Fourier modes of the velocity evolve independently from each other and have variance $\|\widehat{\vect{u}}\|^2 \propto \bar{\sigma}_0^2$, independent of the viscosity and the wavevector. Therefore, 
the three-dimensional velocity spectrum scales as $E_k\sim 
\sigma_0^2 k^2$, compatible with thermal equilibrium of the lowest Fourier modes.

By employing the pre-factor $(\Delta \bar{k})^2$ in the definition  \eqref{eq_def_forcing}, $\bar{\sigma}_0$ has the physical units of inverse time and it determines the Kolmogorov time scale of the flow, independently of the domain size $\bar{L}$.
Furthermore, the forcing \eqref{eq_def_forcing} is in good approximation statistically isotropic, and the correlations of its gradients approximate the isotropic correlation of the forcing \eqref{eq_f_correl_phys} employed in our low-Reynolds number model.
By substituting the definition \eqref{eq_def_forcing} into equation \eqref{eq_f_correl_phys}, it follows that the dimensional forcing amplitude $\bar{\sigma}_0$ used in the DNS, and the non-dimensional amplitude $\sigma=1/\sqrt{15}$ employed in the Fokker-Planck equation \eqref{eq_FPE_Cart}, are related through
\begin{align}
\sigma^2 
\simeq 
\frac{\bar{\gamma}_0^2\bar{\tau}_\eta^2}{30\bar{\nu}} \left(\frac{8}{(\Delta\bar{k})^2}\sum_l \bar{k}_l^4 \bar{\nu} \bar{\sigma}_0^2\right)
\simeq 
\frac{16\pi}{105 \Delta \bar{k}^5}  \bar{\gamma}_0^2 \bar{K}^7  \bar{\tau}_\eta^2 \bar{\sigma}_0^2,
\end{align}
where $\bar{\gamma}_0$ is the characteristic scale of the damping \eqref{eq_Lapl_Gauss}, and $\bar{K}$ is the  maximum forced wavenumber.
We compute the DNS parameter $\bar{\sigma}_0$ that gives a unitary Kolmogorov time scale in the following Appendix \ref{app_Wyld}.
The simulation parameters are summarized in table \ref{tab_DNS}.

\begin{table}
  \begin{center}
\def~{\hphantom{0}}
  \begin{tabular}{lcc}
    & \makecell{Low-Reynolds\\ number simulations} & \makecell{Moderate-Reynolds\\ number simulation}  \\
 Resolved Fourier modes & $64^3$ & $512^3$ \\
Reynolds number & $\Rey_\gamma \simeq 0.2\times 10^{n/10}$ & $\Rey_\lambda\simeq 100$ \\
 Viscosity ($\bar{\nu}$) & $10^{-n/10}$ & $0.002$ \\
 Kolmogorov time scale ($\bar{\tau}_\eta$) & $1$ & $0.09$\\
 Forcing amplitude ($\bar{\sigma}_0$) & $0.027$  & $0.3$ \\
 Maximum forced wavenumber ($\bar{K}$) & $\sqrt{7}$ & $\sqrt{7}$
\end{tabular}
  \caption{Simulation parameters in code units, for the DNS at low and moderate Reynolds numbers. In the low-Reynolds number simulations, $n$ is an integer between 1 and 20, that regulates the viscosity and thus $\Rey_\gamma$.}
\label{tab_DNS}
\end{center}
\end{table}

\begin{figure}
\centerline{
\begin{overpic}[width=\textwidth]{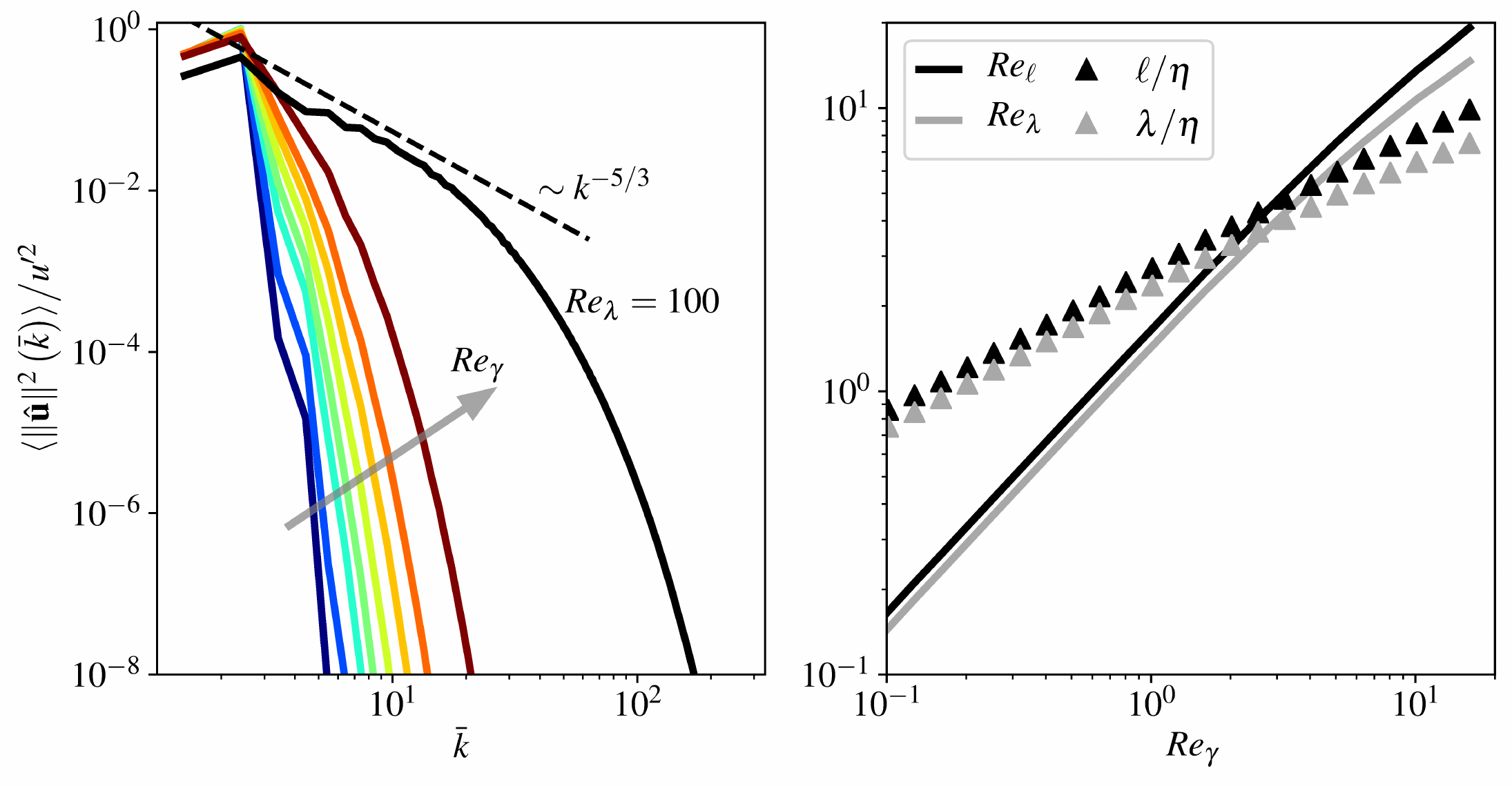}
\put(28,51){(a)}
\put(78,51){(b)}
\end{overpic}
}
	\caption{(a) Kinetic energy spectra from DNS at various Reynolds numbers, with $\Rey_\gamma$ ranging from $0.1$ to $10$ (same color scheme of figure \ref{fig_correlations}). (b) Integral-scale Reynolds number $\Rey_\ell$ and Taylor Reynolds number $\Rey_\lambda$, together with the scale separations $\ell/\eta$ and $\lambda/\eta$, as functions of the low-Reynolds number perturbation parameter $\Rey_\gamma$.}
\label{fig_DNS_setup}
\end{figure}

To get an idea of the parameter range and flow configuration, we report in figure \ref{fig_DNS_setup} the velocity spectra together with a few characteristic Reynolds numbers and scale ratios.
Figure \ref{fig_DNS_setup}(a) shows the kinetic energy spectra from the simulations at various Reynolds numbers. As the Reynolds number increases, energy cascades towards the small scales and the high-wavenumber modes become more energetic, while low-wavenumber modes lose energy. The characteristic inertial range trend $k^{-5/3}$ is noticeable only at the largest Reynolds. Indeed, the phenomenological arguments yielding that inertial range slope require a well-defined scale separations in the flow, which are absent in random low-Reynolds number flows.

Figure \ref{fig_DNS_setup}(b) shows a comparison between the various Reynolds numbers and the corresponding scale separation.
The integral-scale Reynolds number $\Rey_\ell=u'\ell/\nu$ (where $u'=\sqrt{2\int\de k E_k/3}$ is the root mean square velocity and $\ell=\pi\int\de k E_k/(2k u'^2)$ is the integral scale) is proportional to $\Rey_\gamma$ at low Reynolds numbers, while it starts following a different power law at $\Rey_\gamma\simeq 5$, when the transition to turbulence takes place.
The Taylor Reynolds number $\Rey_\lambda=u'\lambda/\nu$  (where $\lambda=\sqrt{15} u'\tau_\eta$ is the Taylor micro-scale) is always smaller than $\Rey_\ell$, it is linear in $\Rey_\gamma$ at low Reynolds numbers while it grows slower starting at $\Rey_\gamma\approx 5$.
We quantify the scale separation through $\ell/\eta$ and $\lambda/\eta=\sqrt{15}u'/u_\eta$, where $u_\eta=\eta/\tau_\eta$ is the Kolmogorov velocity scale. 
While the $\ell$ and $\lambda$ are larger than the Kolmogorov scale already at very small $\Rey_\gamma$, they grow only as $\sqrt{\Rey_\gamma}$ at small Reynolds numbers, resulting in very moderate scale separation at $\Rey_\gamma\gtrapprox 1$. At such Reynolds numbers however, the skewness and preferential alignments in the velocity gradient statistics are already non-negligible, while signatures of intermittency start appearing.
This is consistent with the observation that non-Gaussianity can be pronounced even in flows which do not feature any scale separation \citep{Schumacher2007,Yakhot2018}.

\section{Determining the model parameters through the Wyld expansion}
\label{app_Wyld}

In this Appendix, we compute the low-Reynolds number model parameters, namely the reference length $\bar{\gamma}_0$, the forcing amplitude $\bar{\sigma}_0$, together with the growth rates of the strain self amplification $\avg{\inv_3}$ and of the fourth order moment $\avg{\inv_5}$,
\begin{align}
S_3 = \left.\frac{\partial \avg{\inv_3}}{\partial\Rey_\gamma}\right|_{\Rey_\gamma=0},
&&
X_5 = \left.\frac{1}{2}\frac{\partial^2 \avg{\inv_5}}{\partial\Rey_\gamma^2}\right|_{\Rey_\gamma=0}.
\label{eq_def_growth_rates}
\end{align}
We derive those quantities from the Navier-Stokes equations, by using the weak-coupling expansion proposed by \cite{Wyld1961} for a generic forcing. Although divergent, this power series in the perturbation parameter $\Rey_\gamma$ is useful to derive the moments of the velocity gradient at very small Reynolds numbers, that is the range we are interested in.
We specialize this weak-coupling expansion for a Gaussian, white in time, forcing, and with the aid of symbolic calculus \citep{Sympy2017}, we get the velocity gradient moments up to second order in $\Rey_\gamma$. The computation quickly becomes very involved as the order of the expansion in the Reynolds  number increases, and we illustrate the procedure for the coefficient $S_3$, encoding a first-order correction in $\Rey_\gamma$.

\subsection{Setup for the Wyld expansion}

We consider the space-time Fourier transform of the velocity field $\vect{u}(\vx,t)$,
\begin{align}
\Fu_i(\Fk) = 
\frac{1}{TL^3}
\int\de t \de\vx u_i (\vx,t)\exp(\imag\omega t - \imag\vk\cdot\vx),
\end{align}
where  $\Fk=(\omega,\vk)$ is the position vector in the frequency-wavevector space, and $T$ and $L$ are the periods of the velocity field in time and space respectively.
Following \cite{Wyld1961}, we formally expand the velocity field in powers of the Reynolds number
\begin{align}
    \Fu_i(\Fk)=\sum_N \Fu^{(N)}_i(\Fk) \Rey_\gamma^N,
    \label{eq_u_expansion}
\end{align}
with integer $N\ge 0$. The coefficients $\tilde{\vect{u}}^{(N)}$ at various orders in $\Rey_\gamma$ are related through the Navier-Stokes equations \eqref{eq_NS_spectral} in non-dimensional form
\begin{align}
\left(-\imag\omega + k^2\right) \Fu_i + \imag \Rey_\gamma P_{ij} k_l \left(\Fu_j*\Fu_l\right)
=
\sigma_0 k \tilde{F}_i,
\label{eq_NS_Fourier}
\end{align}
where the non-dimensional parameter $\sigma_0$ relates to the corresponding dimensional (DNS) parameter in equation \eqref{eq_NS_spectral}, $\sigma_0 = \bar{\tau}_\eta\bar{\sigma}_0$, and the symbol $*$ denotes convolution,
\begin{align}
\left[\Fu_j*\Fu_l\right](\Fk) = \frac{TL^3}{(2\pi)^4}\int\de \Fk' \Fu_j(\Fk') \Fu_l(\Fk-\Fk').
\end{align}
Plugging the series expansion of the velocity field \eqref{eq_u_expansion} into the Navier-Stokes equations \eqref{eq_NS_Fourier} gives the following relations for the series coefficients $\vect{\Fu}^{(N)}$,
\begin{subequations}
\begin{align}
\Fu^{(0)}_i(\Fk) &= \sigma_0 k G(\Fk)\tilde{F}_i \label{eq_FNS_0}
\\
\Fu^{(N+1)}_i(\Fk) &= G_{ijl}(\Fk) \sum_{L+M=N} 
\left[\Fu^{(L)}_j * \Fu^{(M)}_l\right](\Fk),
\label{eq_FNS_1}
\end{align}
\label{eq_FNS}
\end{subequations}
where $L,M,N$ are non-negative integers and the propagators read
\begin{align}
G(\Fk) = \frac{1}{-\imag\omega+k^2}, && 
G_{ijl}(\Fk) = -\frac{\imag}{2}  G(\Fk) \left(P_{il}k_j + P_{ij}k_l\right).
\end{align}
The external Gaussian forcing is delta-correlated in time, with zero mean and correlation
\begin{align}
\avg{\tilde{F}_i(\Fk)\tilde{F}_j(\Fk')} =
2\frac{(2\pi)^5}{T^2 L^4}
\sum_l
 P_{ij}(\vect{k}_l)
\left(\delta(\vect{k}-\vect{k}_l) + \delta(\vect{k}+\vect{k}_l)\right) 
\delta(\Fk+\Fk')
\label{eq_f_correl}
\end{align}
where $\vect{k}=\bar{\gamma}_0\bar{\vect{k}}$, the wavevectors $\bar{\vect{k}}$ have integer components, and the forced wavevectors belong to the interval $\{ 1\le \|\bar{\vect{k}}_l\|^2< 7 \}$ (cf.\ DNS setup in Appendix \ref{app_DNS_details}).
The discrete forcing \eqref{eq_f_correl} is the Fourier transform with respect to time of the forcing  employed in the DNS \eqref{eq_def_forcing}, in which the wavevectors are discrete by construction.
This discrete forcing $\tilde{\vect{F}}$ is only approximately isotropic, and the spatial correlation of the forcing employed in the Wyld expansion and in the DNS \eqref{eq_f_correl} corresponds only approximately to the isotropic form \eqref{eq_f_correl_phys} used in our low-Reynolds number model. However, for the chosen forced wavenumbers range, 
the discrepancies between the actual forcing correlation and its isotropic limit are tiny.

\subsection{Moments of the velocity gradient at order zero in the Reynolds number}

We are interested in the velocity gradients moments resulting from the expanded equations \eqref{eq_FNS}, which can be derived from the Fourier-space velocity field as follows. The Fourier transform of the strain rate is a linear operator on the velocity field, and at any order in the Reynolds number we have
\begin{subequations}
\begin{align}
\FS^{(N)}_{ij} = \FS_{ijl} \Fu^{(N)}_{l}, &&
\FS_{ijl} = \frac{\imag}{2} \left( P_{il}k_j + P_{jl}k_i \right).
\end{align}
\label{eq_S3}
\end{subequations}
The zeroth-order strain-rate magnitude follows from equation \eqref{eq_FNS_0} combined with the definition of the forcing correlation \eqref{eq_f_correl},
\begin{align}
\avg{\inv_1} = \avg{S_{ij}^{(0)}S_{ji}^{(0)}} =
\frac{V}{2} \int\de\Fk k_ik_i\avg{\Fu_{j}^{(0)}(\Fk) \Fu_{j}^{(0)}(-\Fk)} = 684\sigma_0^2,
\label{avg_S2}
\end{align}
where $V=TL^3/(2\pi)^4$ is the convolution normalization factor.
The result \eqref{avg_S2} is valid not only at order zero, but for all Reynolds numbers since the random forcing $\tilde{\vect{F}}$ produces a constant variance of the gradients \citep{Furutsu1963,Novikov1965}, independent of the weight of the nonlinearities in \eqref{eq_NS_ndim}.
Imposing a unitary Kolmogorov time scale, $\avg{\inv_1}=1/2$, yields
\begin{align}
\bar{\sigma}_0 \simeq 0.027,
\end{align}
that is the parameter employed in our DNS (cf.\ table \ref{tab_DNS}).

Finally, we need to compute the damping parameter $\bar{\gamma}_0$ from the dimensional velocity gradient field since it constitutes the reference length in our low-Reynolds number model. This reference length follows from the conditional Laplacian \eqref{eq_Lapl_Gauss} in a Gaussian random field \citep{Wilczek2014},
\begin{align}
\avg{\bar{\nabla}^2 \tens{S}^{(0)}\middle|\tens{S}^{(0)}} = -\frac{\tens{S}^{(0)}}{\bar{\gamma}_0^2}.
\label{eq_Lapl_Gauss}
\end{align}
Taking the double contraction of both sides of equation \eqref{eq_Lapl_Gauss} with the strain rate itself and averaging yields
\begin{align}
\bar{\gamma}_0^2 = \frac{\avg{S_{ij}^{(0)}S_{ij}^{(0)}}}{\avg{\bar{\nabla}_kS_{ij}^{(0)}\bar{\nabla}_k S_{ij}^{(0)}}} = \frac{\int\de\Fk \bar{k}^2\avg{\Fu_{j}^{(0)}(\Fk) \Fu_{j}^{(0)}(-\Fk)}}{\int\de\Fk \bar{k}^4 \avg{\Fu_{j}^{(0)}(\Fk) \Fu_{j}^{(0)}(-\Fk)}} 
\simeq 0.20,
\label{eq_gamma0}
\end{align}
where the wavevector norm $\bar{k}$ and the coefficient $\bar{\gamma}_0$ are dimensional (in code units).
The expansion parameter occurring in the asymptotic solution of the FPE \eqref{eq_FPE_sol} $\Rey_\gamma$, is proportional to $\bar{\gamma}_0^2$, as in \eqref{def_Re_gamma}. The relation between $\Rey_\gamma$ and the Reynolds number based on the Taylor microscale $\Rey_\lambda$, that is usually employed in the literature to investigate the onset of non-Gaussianity \citep{Gotoh2022}, depends on the details of the forcing.

\subsection{Moments of the velocity gradient at first and second order in Reynolds}

We are now left with determining the growth rates of the gradient moments, $S_3$ and $X_5$ \eqref{eq_def_growth_rates}, from the Wyld expansion of the Navier-Stokes equations \citep{Wyld1961,Lvov2023}.
For the third order moment of the strain rate at first order in $\Rey_\gamma$, we have 
\begin{align}
S_3 = 3\avg{S^{(1)}_{i_1j_1}S^{(0)}_{j_1j_2}S^{(0)}_{j_2i_1}} = 3 V^2\int\de\Fk_2\de\Fk_1
&
\FS_{i_1j_1 l_1}(\Fk_1) 
\FS_{j_1j_2 l_2}(\Fk_2-\Fk_1) 
\FS_{j_2i_1 l_3}(-\Fk_2)\times\nonumber\\
\times&\avg{\Fu^{(1)}_{l_1}(\Fk_1) \Fu^{(0)}_{l_2}(\Fk_2-\Fk_1)\Fu^{(0)}_{l_3}(-\Fk_2)}.
\end{align}
The moment splits into a deterministic part and a stochastic part, on which the ensemble average acts. We aim to have only Gaussian variables inside that ensemble average, and to this end we iteratively substitute $\Fu_i^{(N)}$ in terms of lower order velocities, up to $\Fu_i^{(0)}$, using the Navier-Stokes equations \ref{eq_FNS}.
At first order, this substitution yields
\begin{align}
S_3 = 3V^3\int\de\Fk_3\de\Fk_2\de\Fk_1
&
\FS_{i_1j_1 l_5}(\Fk_1) G_{l_5l_1l_4}(\Fk_1) 
\FS_{j_1j_2 l_2}(\Fk_2-\Fk_1) 
\FS_{j_2i_1 l_3}(-\Fk_2)\times\nonumber\\
\times&\avg{\Fu^{(0)}_{l_1}(\Fk_1-\Fk_3) \Fu^{(0)}_{l_4}(\Fk_3) \Fu^{(0)}_{l_2}(\Fk_2-\Fk_1)\Fu^{(0)}_{l_3}(-\Fk_2)}.
\label{eq_S3_det_stc}
\end{align}
Substitution of higher order velocities in terms of the Gaussian random field $\tilde{\vect{u}}^{(0)}$ introduces additional convolutions, and for higher-degree moments this iterative substitution leads to rapid proliferation of terms \citep{Monin1975,Lvov2023}, requiring handling the the symbolic/numeric computation of several Wyld-Dyson integrals.

Since $\tilde{\vect{u}}^{(0)}$ is Gaussian, the average in \eqref{eq_S3_det_stc} splits into pairs \citep{Isserlis1918,Wick1950}
\begin{align}
S_3 =& 3 V^3 \int\de\Fk_3\de\Fk_2\de\Fk_1
\FS_{i_1j_1 l_5}G_{l_5l_1l_4}(\Fk_1) 
\FS_{j_1j_2 l_2}(\Fk_2-\Fk_1) 
\FS_{j_2i_1 l_3}(-\Fk_2)\nonumber\\
\Big(
&\avg{\Fu^{(0)}_{l_1}(\Fk_1-\Fk_3) \Fu^{(0)}_{l_4}(\Fk_3)} \avg{\Fu^{(0)}_{l_2}(\Fk_2-\Fk_1) \Fu^{(0)}_{l_3}(-\Fk_2)} + \nonumber\\
&\avg{\Fu^{(0)}_{l_1}(\Fk_1-\Fk_3) \Fu^{(0)}_{l_2}(\Fk_2-\Fk_1)} \avg{\Fu^{(0)}_{l_4}(\Fk_3) \Fu^{(0)}_{l_3}(-\Fk_2)} + \nonumber\\
&\avg{\Fu^{(0)}_{l_1}(\Fk_1-\Fk_3) \Fu^{(0)}_{l_3}(-\Fk_2)} \avg{ \Fu^{(0)}_{l_4}(\Fk_3) \Fu^{(0)}_{l_2}(\Fk_2-\Fk_1)}
\Big).
\label{eq_S3_Wick}
\end{align}
Moreover, for delta-correlated noise, the average acts just as combinations of contractions on the deterministic part, namely
\begin{align}
&S_3 = 
3V^3\int\de\Fk_3\de\Fk_2
\FS_{i_1j_1 l_5}(\bm{0}) G_{l_5l_1l_1}(\bm{0}) 
\FS_{j_1j_2 l_2}(\Fk_2) 
\FS_{j_2i_1 l_2}(-\Fk_2)
R(\Fk_3)
R(\Fk_2) + \nonumber\\
+&
3V^3\int\de\Fk_2\de\Fk_1
\FS_{i_1j_1 l_5}(\Fk_2-\Fk_1)G_{l_5l_1l_4}(\Fk_2-\Fk_1) 
\FS_{j_1j_2 l_1}(\Fk_1)
\FS_{j_2i_1 l_4}(-\Fk_2)
R(\Fk_1)
R(\Fk_2) + \nonumber\\
+&
3V^3\int\de\Fk_3\de\Fk_2
\FS_{i_1j_1 l_5}(\Fk_2+\Fk_3)G_{l_5l_1l_2}(\Fk_2+\Fk_3) 
\FS_{j_1j_2 l_2}(\Fk_3) 
\FS_{j_2i_1 l_1}(-\Fk_2)
R(\Fk_2)
R(\Fk_3)
\label{eq_S3_semifinal}
\end{align}
where for notation convenience $R=\sigma_0^2k^2|G|^2\avg{\tilde{F}_i(\vect{q})\tilde{F}_i(-\vect{q})}$ is the auto-correlation of $\tilde{\vect{u}}^{(0)}$.
At higher order, the number of independent contractions, stemming from the factorization of the higher-order Gaussian moments into products of correlations, rapidly increases introducing an additional technical difficulty.

We tackle the frequency-wavenumbers integrals \eqref{eq_S3_semifinal} by means of a combination of symbolic calculus and discrete integration.
The integration with respect to the frequency is carried out analytically by using the residue theorem \citep{Kleinert2001}.
On the other hand, the integration over the wavevectors reduces to a discrete summation over the discrete forced wavevectors, due to the form of the forcing correlation \eqref{eq_f_correl}, consisting of a superposition of Dirac delta functions.
Evaluating the integral \eqref{eq_S3_semifinal}, and the corresponding one for $X_5$, we finally get
\begin{align}
S_3 \simeq -0.027, &&
X_5 \simeq  -0.0010,
\end{align}
matching well the trend of the moments from our DNS at low $\Rey_\gamma$ (cf.\ figure \eqref{fig_mom_invs}).

For the parameters $X_5$ the procedure is the same as sketched for $S_3$, but the expansion goes up to second order in the Reynolds number and there is one additional integration variable. Also, the average of the sixth order moments of $\Fu^{(0)}$ now splits into fifteen pairs by the Wick theorem (differently from equation \eqref{eq_S3_Wick}, in which we got just three pairs).
Therefore, the direct computation of integrals like \eqref{eq_S3_semifinal} presented here is computationally expensive, and the (equivalent) numerical integration of the expanded Navier-Stokes equations \eqref{eq_NS_Fourier} is preferable. This integration yields the fields $\tilde{\vect{u}}^{(N)}$ \eqref{eq_u_expansion}, from which one can compute the moments of the gradient in physical space through Fast Fourier Transform (the FFT requiring just $13^3$ grid points for these low-order moments and the forcing \eqref{eq_f_correl}).
Systematic study of expressions like \eqref{eq_S3_semifinal}, aiming to an efficient symbolic/numeric evaluation of these integrals, is the subject of ongoing work.

\bibliographystyle{jfm}
\bibliography{jfm_small_Re}

\end{document}